\documentclass[aps,pra,twocolumn,superscriptaddress,nofootinbib]{revtex4-1}

\usepackage{amsmath,amssymb,amsfonts}
\usepackage{graphicx}
\usepackage{dcolumn}
\usepackage{mathtools,bm,bbold}
\usepackage{hyperref}
\usepackage[sort&compress]{natbib}
\usepackage{color}
\usepackage[mathscr]{euscript}

\allowdisplaybreaks

\newcommand{\mbf}{\mathbf}
\newcommand{\sgn}{{\text{sgn}}}
\newcommand{\cl}{{\text{cl}}}

\newcommand{\cS}{{\mathscr{S}}}
\newcommand{\cA}{{\mathscr{A}}}
\newcommand{\cM}{{\mathscr{M}}}

\renewcommand{\Re}{{\mathrm{Re}}}
\renewcommand{\Im}{{\mathrm{Im}}}

\newcommand{\ri}{{\mathrm{i}\mkern1mu}}

\DeclareMathOperator{\Tr}{Tr}

\begin{document}

\title{Nonequilibrium dynamics of spin-boson models from phase space methods}

\author{Asier Pi\~neiro Orioli}
\email{pineiroorioli@thphys.uni-heidelberg.de}
\affiliation{Institut f\"ur Theoretische Physik, Universit\"at Heidelberg, Philosophenweg 16, 69120 Heidelberg, Germany}
\author{Arghavan Safavi-Naini}
\email{arsa4885@jila.colorado.edu}
\author{Michael L. Wall}
\thanks{Present address: The Johns Hopkins University Applied Physics Laboratory, Laurel, MD 20723, USA}
\author{Ana Maria Rey}
\affiliation{JILA, NIST and University of Colorado, Boulder, CO}

\begin{abstract}
	An accurate description of the nonequilibrium dynamics of systems with coupled spin and bosonic degrees of freedom remains theoretically challenging, especially for large system sizes and in higher than one dimension. Phase space methods such as the Truncated Wigner Approximation (TWA) have the advantage of being easily scalable and applicable to arbitrary dimensions. In this work we adapt the TWA to generic spin-boson models by making use of recently developed algorithms for discrete phase spaces \cite{PhysRevX.5.011022}. Furthermore we go beyond the standard TWA approximation by applying a scheme based on the Bogoliubov-Born-Green-Kirkwood-Yvon (BBGKY) hierarchy of equations \cite{PhysRevB.93.174302} to our coupled spin-boson model. This allows in principle to study how systematically adding higher order corrections improves the convergence of the method. To test various levels of approximation we study an exactly solvable spin-boson model which is particularly relevant for trapped-ion arrays. Using TWA and its BBGKY extension we accurately reproduce the time evolution of a number of one- and two-point correlation functions in several dimensions and for arbitrary number of bosonic modes.
\end{abstract}

\maketitle


\section{Introduction}
\label{sec:intro}

Coupled spin and bosonic degrees of freedom appear in a variety of  condensed matter and atomic, molecular, and optical (AMO) physics systems. As such, a detailed understanding of their nonequilibrium dynamics can have broad applications. While spin-boson models have been studied extensively in condensed matter physics \cite{CALDEIRA1983374}, AMO systems offer a unique platform where the dynamics of both the spins and the bosons can be studied in a controlled manner. For example, many-body spin-boson models, where many spins couple to a single or many bosonic modes, can be engineered using cold atoms in cavities \cite{PhysRevLett.104.073602,RevModPhys.85.553} or trapped ions \cite{PhysRevLett.82.1971,PhysRevLett.92.207901,PhysRevLett.103.120502,Britton2012,PhysRevA.78.010101}. These realizations provide a great deal of flexibility, from the range of interactions to the dimensionality of the system.

Despite the ubiquity of spin-boson models in nature, efficient computational methods for studying the nonequilibrium dynamics are hard to develop. Theoretical approaches are applicable to specific situations. For instance, when the system features a large separation of scales between spins and bosons, the bosons may be adiabatically eliminated \cite{PhysRevLett.92.207901}, resulting in an effective spin model.
Alternatively, in the presence of permutational symmetry of the density matrix one may use computational methods which take advantage of the reduction of the complexity from exponential to polynomial \cite{PhysRevB.91.035306}. However, for many physically relevant systems these idealized approximations are invalid.

Recently, methods based on Matrix Product States (MPSs) have been successfully applied to systems where the spins and bosons contribute on similar footing in the dynamics~\cite{PhysRevA.94.053637} . These include cases with non-uniform interactions, and could in principle also be applied to systems with additional non-commuting terms like a transverse field \cite{PhysRevA.95.013602}. While MPSs can efficiently treat cases with many spins ($N_s\sim100$) when coupled to only one mode, as soon as the number of relevant boson modes is higher only moderate sizes can be simulated ($N_s\sim N_b \sim 10$). Hence, alternative methods are required to handle large system sizes.

In this work, we study the dynamics of a system of spins-$1/2$ coupled to a set of bosonic modes using phase space methods. These methods are based on phase space descriptions of quantum physics using classical variables \cite{gardiner2004quantum,WallsMil1994}. The most prominent example and the starting point of this paper is the Truncated Wigner Approximation (TWA). Observables in TWA are computed by averaging over classical trajectories, which are obtained by sampling the initial conditions from the Wigner (quasi)probability distribution \cite{Hillery1984,Polkovnikov:2009ys} and evolving them with the classical equations of motion.
TWA has been successfully applied to bosonic systems in fields ranging from quantum optics and cold atoms \cite{Polkovnikov:2009ys,Blakie2008} to cosmology \cite{Khlebnikov1996,Son:1996zs} and quantum field theory \cite{AartsBe2002,BergesGas2007}. In recent years, promising progress has been made as well in treating fermionic \cite{Davidson:2016ddv} and spin models \cite{PhysRevLett.114.045701,PhysRevX.5.011022,schachNJP2015} with this method. 
In particular, it has been shown that using discrete, instead of continuous, Wigner functions for the spins can lead to significant improvements in spin-$1/2$ systems \cite{PhysRevX.5.011022}. Here, we apply TWA to a spin-boson model and make use of this discrete sampling for the spins, thus differing from previous works on spin-boson systems \cite{AltlandGu2008}.

The range of validity of TWA is usually linked to either high occupancies or large macroscopic fields. However, TWA is often found to be a successful description beyond its apparent range of validity, for example, in spin-$1/2$ systems, where genuine quantum effects may become relevant. Being able to compute next-order corrections to TWA becomes then an essential task in order to understand and extend its applicability to such systems.
For this we study recently developed extensions of TWA \cite{PhysRevB.93.174302,KastnerDiss2017}, adapted to our spin-boson model, which allow one to add corrections order by order and hence estimate the error of the approximation. In this framework corrections are added by solving a truncated Bogoliubov-Born-Green-Kirkwood-Yvon (BBGKY) hierarchy of equations.

The computational cost of these phase space methods scales with the number of equations to be solved. For TWA this depends linearly on the total number of spins and bosons. The extra computational cost introduced in BBGKY by solving the dynamics of two-point and higher order correlations instead of single-particle terms translates to a larger number of equations to be solved. Nevertheless, this number still scales polynomially with the system size and hence the computational complexity is dramatically reduced compared to the full quantum calculation. Furthermore, unlike unbiased approaches where the dimensionality of the system is a fundamental limitation, here the dimensionality only enters the computational speed indirectly through the number of spins and bosons that are effectively coupled.

While these methods can be applied to any system with spin and bosonic degrees of freedom, here we study an analytically solvable model relevant for trapped-ion experiments.
Despite its integrability, this model exhibits non-trivial many-body phenomena such as spin-squeezing \cite{Bohnet1297}. Furthermore, it admits an effective description in terms of an Ising model with tunable range interactions, $1/r^\alpha$ with $\alpha\sim 0-3$, allowing one to investigate quantum magnetism. As such, it is an ideal model for benchmarking the current experimental realizations of spin-boson simulators \cite{Britton2012,Richerme2014}, as well as approximate computational methods such as those discussed in this paper.

This work is organized as follows: Section~\ref{sec:model} introduces the spin-boson model and gives exact formulas for the evolution of one- and two-point observables. The application of the TWA method to spins and bosons is presented in Section~\ref{sec:twa} and the first results on the single mode case are discussed in Section~\ref{sec:singlemode}. In Section~\ref{sec:bbgky} we show how corrections may be added to the spin-boson TWA using BBGKY equations and apply this to the many-mode case in Section~\ref{sec:manymode}. Section~\ref{sec:nosampling} discusses the importance of sampling the initial conditions and in Section~\ref{sec:thermal2D} we show the full capabilities of the method by computing the evolution of a many-spin, many-boson, two-dimensional system with experimentally relevant thermal initial conditions for the bosons. Finally, we give an outlook and present our conclusions in Section~\ref{sec:conclusions}.


\section{Spin boson model}
\label{sec:model}

We consider a system of trapped ions in one or two spatial dimensions where the two internal states of the ions, modeled as an effective spin-$1/2$, are coupled to the phonon modes of the ion crystal~\cite{Britton2012,PhysRevLett.103.120502}. The ion crystal is formed due to the interplay of the Coulomb repulsion between the ions and the external electromagnetic trapping potentials, and supports a set of normal modes. The spin-phonon coupling is generated by lasers in a Raman scheme used to impart net momentum $k_{R}$ along the direction perpendicular to the crystal structure~\cite{leibfried2003experimental}.  Following a frame transformation on the spins, the Hamiltonian can be expanded to the lowest order in the Lamb-Dicke parameter $\eta_{\mu}=k_R\sqrt{1/2M\omega_{\mu}}$, where $M$ is the mass of an ion, $\{\omega_{\mu}\}$ are the phonon normal mode frequencies, and $\hbar=k_B=1$ unless otherwise specified. The Hamiltonian takes the form
\begin{equation}
\hat H= \hat H_{\rm boson}+\hat H_{\rm s-b}
\end{equation}
with
\begin{align}
\hat{H}_{\mathrm{boson}} &=\sum_{\mu=1}^{N_b}\omega_{\mu} \hat{n}_{\mu},\\
\hat{H}_{\mathrm{s-b}} &=-F \cos\left(\omega_R t\right) \sum_{i=1}^{N_s} \sum_{\mu=1}^{N_b} b_{i\mu}\frac{\eta_{\mu}}{k_R}(\hat{a}_{\mu}+\hat{a}_{\mu}^{\dagger})\hat{\sigma}^z_i\, .
\label{eq:Hions}
\end{align}
Here, $\hat{n}_\mu=\hat{a}^\dagger_\mu \hat{a}_\mu$, $\omega_R$ is the Raman beatnote frequency of the beams which create a spin dependent force with magnitude $F$, and  $b_{i \mu}$ is the amplitude of the vibrational mode $\mu$ at site $i$ in units of the normal mode oscillator length.

If $\omega_R \approx \omega_\mu$, for some mode $\mu$, one can perform the Rotating Wave Approximation (RWA), following which we obtain the spin-boson Hamiltonian that will be used throughout this work, namely
\begin{align}
	\hat H_{\rm RWA}=&\, - \frac{1}{2} \sum_{i,\mu} \Omega_{i\mu} \left( \hat{a}_\mu e^{\ri\delta_\mu t} + \hat{a}^\dagger_\mu e^{-\ri\delta_\mu t} \right) \hat{\sigma}^z_i \, .
\label{eq:HRWA}
\end{align}
Here, $\Omega_{i\mu} \equiv \Omega_\mu b_{i\mu} = F b_{i \mu}\eta_\mu/k_R$ and $\delta_\mu = \omega_R - \omega_\mu$ is the detuning from the near phonon mode. For small $\Omega_{i\mu}/\delta_\mu$ or alternatively at stroboscopic times given by $t=2\pi n/\delta_\mu$ ($n\in\mathbb{Z}$) the dynamics of the spins effectively decouples from the bosons and is described by an effective Ising model given by 
\begin{equation}
	\hat{H}_{\textrm{Ising}} = \sum_{i<j} J_{ij} \hat \sigma^z_i \hat \sigma^z_j \, ,
\label{eq:isingH}
\end{equation}
where $J_{ij}=1/2\sum_\mu \Omega_{i\mu}\Omega_{j\mu} /\delta_\mu$~\cite{KimChang2009}. For positive detunings the spin-spin coupling is well approximated by $J_{ij}\sim1/r_{ij}^\alpha$, where $\alpha\in[0,3)$ can be tuned by $\omega_R$. Even outside its strict validity range this mapping to a spin-only Hamiltonian is useful to understand the dynamics of the full spin-boson model.

We use the model of Eq.~(\ref{eq:HRWA}) to benchmark the accuracy of our method by comparing the dynamics of different observables to their exact forms. For most of the work we let the system start from an initial state ${\vert \psi(t=0)\rangle}={\vert \rightarrow \rangle^{\otimes N_s}} \otimes {\vert 0 \rangle^{\otimes N_b}}$, where $\vert \rightarrow \rangle$ denotes a spin pointing along the $+ \hat x$ direction and $\vert 0 \rangle$ is the vacuum state of the phonons. However, we will consider as well more general initial conditions, such as rotated states $(\cos(\theta/2) {|\!\! \uparrow\rangle} + \sin(\theta/2) |\!\! \downarrow\rangle)^{\otimes N_s}$ with $\theta\neq\pi/2$ for the spins (see Section~\ref{sec:nosampling}). Since trapped ion experiments operate at finite temperatures we consider as well the more general case of a thermal initial state for the phonons given by $\hat{\rho}_{th}\equiv \bigotimes_\mu e^{-\beta_\mu \hat{H}_{b,\mu}}/\Tr(e^{-\beta_\mu \hat{H}_{b,\mu}})$ with $\hat{H}_{b,\mu}=\omega_\mu \hat{n}_\mu$ and a mode-dependent inverse temperature $\beta_\mu=1/T_\mu$ (see Section~\ref{sec:thermal2D}).

We compute the ab-initio dynamics of three sets of observables: the collective spin $\langle \hat{S}_x \rangle \equiv \sum_i \langle \hat{\sigma}_i^x \rangle / N_s$, spin-spin two-point correlators $\langle \hat{\sigma}_i^\alpha \hat{\sigma}_j^\beta \rangle_c \equiv \langle \hat{\sigma}_i^\alpha \hat{\sigma}_j^\beta \rangle - \langle \hat{\sigma}_i^\alpha \rangle \langle \hat{\sigma}_j^\beta \rangle$ and spin-phonon correlators of the form $\langle \hat{a}_\mu \hat{\sigma}_i^\alpha \rangle_c \equiv \langle \hat{a}_\mu \hat{\sigma}_i^\alpha \rangle - \langle \hat{a}_\mu \rangle \langle \hat{\sigma}_i^\alpha \rangle$. For later comparison with the proposed method we give the exact formulas for the evolution of these observables under the Hamiltonian $\hat{H}_{\rm RWA}$ for initial states with the spins pointing in $+\hat{x}$ and the phonons thermally occupied. We begin by measuring the site-resolved magnetization $\langle \hat{\sigma}_i^x\rangle$ given by \cite{PhysRevA.93.013415}
\begin{align}
	\langle \hat{\sigma}_i^x \rangle =&\, e^{-\Gamma_i(t)} \prod_{k\neq i} \cos\left( \varphi_{ik}(t) \right) \, ,
\label{eq:expX_quantum}
\end{align}
where
\begin{align}
	\Gamma_i (t) \equiv&\, 4 \sum_\mu \left( \bar{n}_\mu + \frac{1}{2} \right) \frac{\Omega_{i\mu}^2}{\delta_\mu^2} \sin^2(\delta_\mu t/2) \, ,
\label{eq:def_Gamma}\\
	\varphi_{ij} (t) \equiv&\, \sum_\mu \frac{\Omega_{i\mu}\Omega_{j\mu}}{\delta_\mu} \left( t - \frac{\sin(\delta_\mu t)}{\delta_\mu} \right) \, ,
\label{eq:def_varphi}
\end{align}
and
\begin{equation}
	\bar n_\mu \equiv \langle \hat{n}_\mu \rangle (t=0) = \frac{1}{ e^{-\beta_\mu \omega_\mu} -1 }  \, .
\label{eq:nbar}
\end{equation}

The product of cosine functions in Eq.~\eqref{eq:expX_quantum} describes the depolarization of the collective spin, while the factor $e^{-\Gamma_i(t)}$ adds oscillations on top of it. This dynamics can also be understood if one studies the effective spin model (\ref{eq:isingH}), where the coherent depolarization of the collective spin length is caused by the Ising coupling. In this case, one obtains the same evolution as in (\ref{eq:expX_quantum}) but with $\Gamma_i\equiv0$ and $\varphi_{ij}(t)\rightarrow 2 J_{ij} t$. In fact, the latter substitution becomes exact in the long time limit, as it can be seen from the full expression (\ref{eq:def_varphi}) for $\varphi_{ij}(t)$.

The interactions between the spins lead as well to the buildup of spin-spin correlations, which evolve according to
\begin{align}
	\langle \hat{\sigma}_i^x \hat{\sigma}_j^x \rangle =&\, \frac{1}{2}\left\{ e^{-\Gamma_{ij}^-} \prod_{k\neq i,j} \cos\left( \varphi_{ik} - \varphi_{jk} \right) \right. \nonumber\\
	&\, + \left. e^{-\Gamma_{ij}^+} \prod_{k\neq i,j} \cos\left( \varphi_{ik} + \varphi_{jk} \right) \right\} \, ,
\label{eq:expXX_quantum}
\end{align}
where
\begin{align}
	\Gamma_{ij}^\pm \equiv&\, 4 \sum_\mu \left( \bar{n}_\mu + \frac{1}{2} \right) \frac{\left( \Omega_{i\mu} \pm \Omega_{j\mu} \right)^2}{\delta_\mu^2} \sin^2 (\delta_\mu t/2) \, .
\end{align}
The evolution of the $yy$-component shows a similar behavior to (\ref{eq:expXX_quantum}) and therefore we restrict the discussion to the $xx$-component.
At the same time, the spin-dependent displacement of the phonon modes leads to the growth of spin-phonon correlations as characterized by
\begin{align}
	\langle \hat{a}_\mu \hat{\sigma}_i^y \rangle &= \frac{\ri}{2\delta_\mu}\left( 1-e^{-\ri\delta_\mu t} \right) \, e^{-\Gamma_i(t)}\, \prod_{k\neq i} \cos\left( \varphi_{ik} \right) \nonumber\\
	&\, \times \left\{ 2\left( \bar{n}_\mu + \frac{1}{2} \right) \Omega_{i\mu} - \ri \sum_{m\neq i} \Omega_{m\mu} \tan\left( \varphi_{im} \right) \right\} \, ,
\label{eq:expASy_quantum}
\end{align}
where we note that for our particular initial state $\langle \hat{a}_\mu \hat{\sigma}_i^x \rangle=0$.
The spin-phonon correlations oscillate rapidly with $\delta_\mu$ and are responsible for the $e^{-\Gamma_i}$ and $e^{-\Gamma_i^\pm}$, which imprint oscillations on top of the general trend of the collective spin and the spin-spin correlators, respectively.
Additionally we see that Eq.~(\ref{eq:expASy_quantum}) is proportional to Eq.~(\ref{eq:expX_quantum}) and hence the decay of the magnetization runs parallel to that of the spin-phonon correlator. Further insight into the evolution of these observables will be given in the next sections by numerically evaluating the above expressions.

For the most part of the remainder of this work we consider a system of ions in a linear 1D trap but emphasize that our conclusions are largely independent of the dimensionality of the system and give a 2D example in Section~\ref{sec:thermal2D}.


\section{Truncated Wigner Approximation for spins and bosons\label{sec:twa}}

In the following we use the well-known Truncated Wigner Approximation (TWA) and its recently developed discrete extension \cite{PhysRevX.5.011022} to model the dynamics of the spin-boson system described by Eq.~(\ref{eq:HRWA}).
We begin by a brief description of the method and refer to \cite{Polkovnikov:2009ys} for more details. Consider a single spin $\hat{\boldsymbol{\sigma}}=(\hat{\sigma}^x,\hat{\sigma}^y,\hat{\sigma}^z)$ coupled to a single mode $\hat{a}$.
Within the TWA each spin is replaced by a classical vector $\hat{\boldsymbol{\sigma}} \rightarrow \mbf S=(S^x,S^y,S^z)^T$ and each boson by a complex number $\hat{a} \rightarrow A$ and $\hat{a}^\dagger \rightarrow A^*$.
Similarly, each operator $\hat{O}$ is replaced by a classical function $O_W(\mbf S,A,A^*)$ called the Weyl symbol of $\hat{O}$.
From here on we suppress the dependence on $A^*$ and write $O_W(\mbf S,A)$ instead.
This mapping is formally accomplished using the Wigner-Weyl representation. For bosons it is given by
\begin{equation}
	O_W (A) = \int \mathrm{d}^2 \eta \langle A-\eta/2 | \hat{O} | A+\eta/2 \rangle e^{(\eta^*A-\eta A^*)/2} \, ,
\label{eq:wigner_def}
\end{equation}
where $| \eta \rangle$, $\eta\in \mathbb{C}$, is a coherent state and $\mathrm{d}^2\eta = \mathrm{d}\Re\, \eta \, \mathrm{d}\Im\, \eta/\pi$.
For spins the definition goes along the same lines if one uses a Schwinger boson representation \cite{Polkovnikov:2009ys}.

The Weyl symbol of the density matrix $\hat{\rho}$ is the Wigner function $W$ \cite{Hillery1984,Polkovnikov:2009ys}. This quantity plays the role of a (quasi) probability distribution, in the sense that expectation values of observables are obtained via $\langle \hat{O} \rangle = \int \mathrm{d}^2A\, W(A)\, O_W\big(A\big)$, and an equivalent expression for spin operators.
In the TWA the time evolution of the Wigner function is assumed to be approximately stationary along the classical trajectories. Within this approximation expectation values are hence obtained as 
\begin{align}
	\langle \hat{O}(t;\hat{\boldsymbol{\sigma}},\hat{a}) \rangle \approx&\, \int \mathrm{d}\mbf S_0\, \mathrm{d}^2A_0\, W(\mbf S_0,A_0)\, O_W\big( \mbf S_\cl(t),A_\cl(t) \big) \nonumber\\
	\equiv&\, \langle O_W(t;\mbf S,A) \rangle_\cl \, ,
\label{eq:twa_spinboson}
\end{align}
where $\mathrm{d}\mbf S_0 = \mathrm{d}S_0^x\, \mathrm{d}S_0^y\, \mathrm{d}S_0^z$, $\mathrm{d}^2A_0 = \mathrm{d}\Re A_0 \, \mathrm{d}\Im A_0/\pi$, $A_0\equiv A(0)$, $\mbf S_0\equiv \mbf S(0)$ and $\mbf S_\cl(t)\equiv \mbf S_\cl(t;\mbf S_0,A_0)$ and $A_\cl(t)\equiv A_\cl(t;\mbf S_0,A_0)$ fulfil the classical equations of motion.
Numerically, this expression can be evaluated in a three-step process: Monte Carlo sampling of the initial conditions, evolution with the classical equations of motion and averaging.

The classical variables are initialized as random numbers drawn from the Wigner distribution $W(\mbf S,A)$. One can write this as $\mbf S(0)=\langle \hat{\boldsymbol{\sigma}}(0) \rangle + \delta \mbf S_0$ and $A(0) = \langle \hat{a}(0) \rangle + \delta a_0$, where $\delta \mbf S_0$ and $\delta a_0$ are random noise terms with zero mean. The variance of the noise ensures that the classical system reproduces the initial quantum and statistical fluctuations of the original one.
Each of these initial configurations is then evolved using the classical equations of motion. These can be obtained from the mean-field equations for the expectation values $\langle \hat{\boldsymbol{\sigma}} \rangle$ and $\langle \hat{a} \rangle$ by replacing $\langle \hat{\boldsymbol{\sigma}} \rangle \rightarrow \mbf S$ and $\langle \hat{a} \rangle \rightarrow A$, assuming all products of operators have been previously symmetrized (see e.g.~Eq.~(\ref{eq:symmet})).
After this procedure an expectation value $\langle \hat{O}(\hat{\boldsymbol{\sigma}},\hat{a}) \rangle$ at time $t$ can be computed by averaging the function $O_W(\mbf S_\cl(t),A_\cl(t))$ over all classical trajectories.
For totally symmetrized products of the operators $\hat{\sigma}^\alpha$, $\hat{a}$ and $\hat{a}^\dagger$, the Weyl symbol is given by substituting directly operators with classical variables. For instance,
\begin{equation} 
\label{eq:symmet}
\langle |A|^2 \rangle_\cl \approx \langle \frac{1}{2}(\hat{a} \hat{a}^\dagger+\hat{a}^\dagger \hat{a}) \rangle \, .
\end{equation}
It is important to note that while TWA uses mean-field type equations, it goes well beyond a mean-field approximation. In fact, the averaging over a noise distribution involves taking processes of arbitrary order in fluctuations around mean-field into account.

In this work we are interested in modelling the dynamics of many spins $\mbf S_i$ and many bosons $A_\mu$. To accomplish that we consider only initial states where all the bosons and the spins are uncorrelated. Hence, the Wigner function factorizes as $W(\{\mbf{S}_i\},\{A_\mu\})=\prod_{i} W_s(\mbf{S}_i) \prod_\mu W_b(A_\mu)$ and spins and bosons can be sampled independently from each other. For bosons, we will consider vacuum and free thermal states. In this case, the Wigner function is just given by a Gaussian distribution
\begin{equation}
	W_b(A_{\mu,0})= \frac{1}{2\pi\sigma_\mu^2} \exp\left(-|A_{\mu,0}|^2 / (2\sigma_\mu^2)\right) \, ,
\label{eq:boson_samplingW}
\end{equation}
where $\sigma_\mu^2=(\bar{n}_\mu+1/2)/2$ and $\bar{n}_\mu=0$ for vacuum.
For spins, the statistics of typical states like $|\!\! \uparrow \rangle$ are highly nontrivial and often Gaussian approximations are used that reproduce only the low-order cumulants of the state \cite{Polkovnikov:2009ys,PhysRevLett.114.045701}. However, new sampling schemes based on discrete phase spaces have been recently proposed \cite{WOOTTERS19871,PhysRevX.5.011022} which can capture expectation values to infinite order (see Appendix~\ref{app:IC_TWA}) and which have been proven to outperform continuous approximations. For example, for the $| \uparrow \rangle$ state the Wigner distribution is given by
\begin{equation}
	W_s(\mbf{S}_{i,0})=\frac{1}{4}\,\delta(S^z_{i,0}-1) \sum_{x_i,y_i=\pm1} \delta(S^x_{i,0}-x_i) \delta(S^y_{i,0}-y_i) \, ,
\label{eq:spin_samplingW}
\end{equation}
i.e.~the $z$-component is fixed at $+1$ and the transverse directions are sampled with $\pm1$. Based on this, one can also construct the Wigner distribution for rotated initial conditions, see Appendix~\ref{app:IC_TWA}.

The classical equations that govern the evolution of the classical variables $\{A_\mu\}$ and $\{\mbf{S}_i\}$ can be derived in the following way. First, one derives the Heisenberg equations of motion for the operators $\hat{a}_\mu$ and $\hat{\boldsymbol{\sigma}}_i$. After that, all products of operators need to be totally symmetrized using the corresponding equal-time commutation relations. In particular, products like $(\hat{\sigma}_i^x)^2$ need to be simplified to $\mathbb{1}$. Finally, one simply substitutes $\hat{a}_\mu \rightarrow A_\mu$ and $\hat{\boldsymbol{\sigma}}_i \rightarrow \mbf S_i$. Following this recipe one obtains for the Hamiltonian (\ref{eq:HRWA}) the classical equations of motion
\begin{align}
\begin{aligned}
	\dot{A}_\mu =&\, \frac{\ri}{2} e^{-\ri\delta_\mu t} \sum_i \Omega_{i\mu} S_i^z \, ,\\
	\dot{S}_i^x =&\, 2 \sum_\mu \Omega_{i\mu}\, \Re\left[ ( S_i^y A_\mu) e^{\ri\delta_\mu t} \right] \, , \\
	\dot{S}_i^y =&\, -2 \sum_\mu \Omega_{i\mu}\, \Re\left[ ( S_i^x A_\mu) e^{\ri\delta_\mu t} \right] \, , \\
	\dot{S}_i^z =&\, 0 \, .
\label{eq:ceom_gen}
\end{aligned}
\end{align}


\section{Single mode\label{sec:singlemode}}

\begin{figure*}[t!]
\includegraphics[width=.8\textwidth]{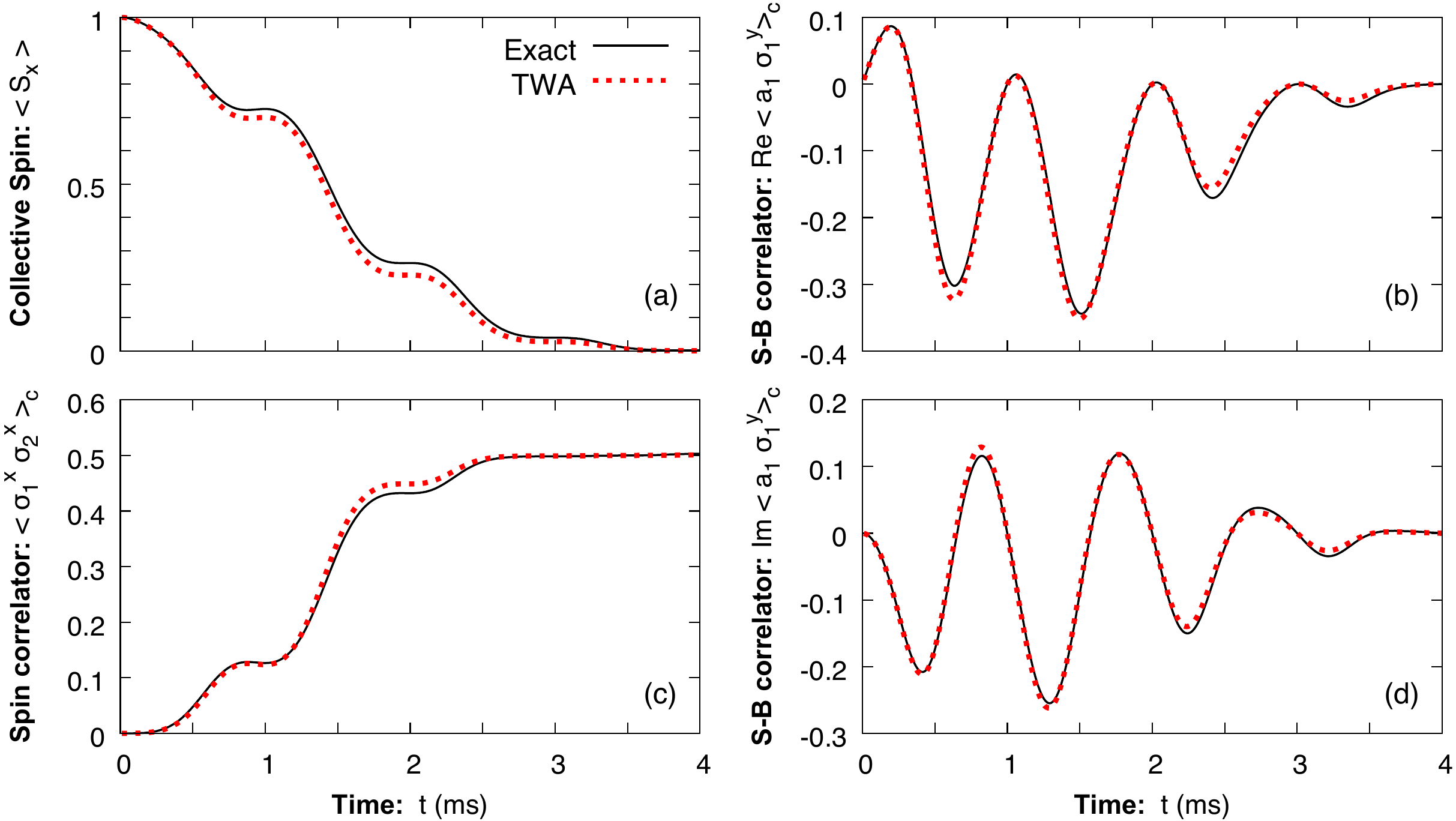}
\caption{Comparison of different observables using TWA and the exact solution for the single mode case in 1D and for $N_s=10$ spins, $\delta=1\,\text{kHz}$ and $\Omega=0.65 \,\text{kHz}$. Panels (a) and (c) show the total magnetization and a spin-spin correlator, whereas panels (b) and (d) show the real and imaginary parts of the two-point correlator between a single spin and the bosonic mode, respectively. The deviation of the TWA solution from the exact behavior is due to the self-interaction terms.}
\label{fig:monroe_single}
\end{figure*}

In this section we apply the TWA to the regime where $\delta_\mu \gg \Omega_\mu$ for all modes except for the center of mass mode (COM) where $\delta_{\mathrm{COM}} \gtrsim \Omega_{\mathrm{COM}}$. In this regime the dynamics of the system is dominated by the COM mode and the homogeneity of this mode generates an effective Ising model with uniform, all-to-all interactions.

Throughout this work we express frequencies in units of $2\pi$ and define for convenience $\delta\equiv\delta_{\mathrm{COM}}$, $\Omega\equiv\Omega_{\mathrm{COM}}$ and $\omega\equiv\omega_{\mathrm{COM}}$.
We consider a chain of $N_s=10$ atoms with parameters $\delta=1\,\text{kHz}$ and $\Omega=0.65 \,\text{kHz}$. To show that for this choice of parameters the influence of the other modes is suppressed we first note that the detuning with respect to other modes can be expressed in terms of the COM frequency as $\delta_\mu=\delta+\Delta\omega_\mu$ where $\Delta\omega_\mu=\omega-\omega_\mu$. The frequency difference with respect to the COM mode ($\mu\neq \mathrm{COM}$) is $\Delta\omega_\mu \geq 80 \,\text{kHz}$, see Appendix~\ref{app:modes}. Similarly one can write $\Omega_\mu = \Omega \sqrt{\omega/\omega_\mu}$.
As a result of this we have $\Omega/\delta=0.65$ and $\Omega_\mu/\delta_\mu \leq 0.008 \ll \Omega/\delta$ for the remaining modes.

We let the system start in a product state of both bosons and spins. The bosons start in the vacuum state $| 0 \rangle^{\otimes N_b}$ with $N_b=1$ and the spins begin in a product state $|\!\! \rightarrow \rangle^{\otimes N_s}$ pointing in the $+x$ direction.
Fig.~\ref{fig:monroe_single}(a) shows the results obtained with TWA for the total magnetization $\langle \hat{S}_x \rangle \equiv \sum_i \langle \hat{\sigma}_i^x \rangle / N_s$ compared to the exact solution. We find a remarkable agreement with the exact solution except for a small deviation. As anticipated, the total magnetization decays to zero due to the phonon-mediated spin-spin interactions while the coupling to the phonons induces oscillations on top at a frequency $\delta$.

The spin-spin interactions create correlations between spins at arbitrary distances. This is quantified by the spin-spin connected two-point correlators $\langle \hat{\sigma}_i^\alpha \hat{\sigma}_j^\beta \rangle_c$.
As Fig.~\ref{fig:monroe_single}(c) shows for $i=1$ and $j=2$, correlations are built up between spins during the time that the total magnetization decreases. In fact, due to the uniformity of the interactions all spin-spin correlators ($i\neq j$) show exactly the same behavior as Fig.~\ref{fig:monroe_single}(c). The correlations between phonon and spins are also the same for all spins. Figs.~\ref{fig:monroe_single}(b) and (d) show the real and imaginary parts of the connected spin-phonon correlator $\langle \hat{a}_1 \hat{\sigma}_1^y \rangle_c$. This quantity oscillates with the frequency $\delta$, while its envelope first grows and then decays to zero as the total magnetization vanishes. Except for small deviations the agreement between TWA and the exact solution is remarkable for both spin-spin and spin-phonon correlation functions.

To understand the origin of the small discrepancies observed in Fig.~\ref{fig:monroe_single} we solve the classical equations of motion (\ref{eq:ceom_gen}) and perform the sampling analytically, see Appendix~\ref{app:exact_twa}. This yields
\begin{align}
	\langle S_i^x \rangle_{\cl} =&\, e^{-\Gamma_i(t)} \prod_k \cos\left( \varphi_{ik}(t) \right) \, ,
\label{eq:expX_twa} \\
	\langle S_i^x S_j^x \rangle_{\cl} =&\, \frac{1}{2}\left\{ e^{-\Gamma_{ij}^-} \prod_{k} \cos\left( \varphi_{ik} - \varphi_{jk} \right) \right. \nonumber\\
	&\,\quad + \left. e^{-\Gamma_{ij}^+} \prod_{k} \cos\left( \varphi_{ik} + \varphi_{jk} \right) \right\} \, ,
\label{eq:expXX_twa} \\
	\langle A_\mu S_i^y \rangle_\cl =&\, \frac{\ri}{2\delta_\mu}\left( 1-e^{-\ri\delta_\mu t} \right) \, e^{-\Gamma_i(t)}\, \prod_k \cos\left( \varphi_{ik} \right) \nonumber\\
	\times& \left\{ 2\left( \bar{n}_\mu + \frac{1}{2} \right) \Omega_{i\mu} - \ri \sum_{m} \Omega_{m\mu} \tan\left( \varphi_{im} \right) \right\} \, .
\label{eq:expASy_twa}
\end{align}
These expressions look identical to the exact quantum solutions, Eqs.~(\ref{eq:expX_quantum}), (\ref{eq:expXX_quantum}) and (\ref{eq:expASy_quantum}), except that the indices $i$ and $j$ are excluded in the sums and products of the quantum solution. This stems from a kind of `self-interaction' that is an artefact of the TWA solution. To see this we compare Eq.~(\ref{eq:expX_twa}) to the expression we obtain by instead integrating out the bosons exactly and then doing TWA. For this we first solve the Heisenberg equation of motion for $\hat{a}_\mu$ at the quantum level and then insert it into the equations for $\hat{\sigma}_i^x$ and $\hat{\sigma}_i^y$.
The $k=i$ in the sum automatically drops out after symmetrization, due to the anti-commutativity of the spin matrices. After symmetrization we substitute operators by classical variables. Due to the excluded $k=i$ term the result of TWA becomes exact, $\langle S^x \rangle_{\cl} = \langle \hat{S}^x \rangle$, for the initial conditions considered here. Similar self-interaction terms in higher order products are also responsible for the differences observed in the two-point correlators, but identifying them becomes more involved.

For a single mode, this self-interaction term leads only to a small discrepancy which, in fact, becomes negligible as the number of spins $N_s$ is increased, see also the results of Section~\ref{sec:thermal2D}. However, the situation can be different when many modes contribute to the dynamics.
To illustrate this we use $\varphi_{ij} \sim \sum_\mu b_{i\mu}b_{j\mu}/(\omega_\mu \delta_\mu)$, valid at times $t\gg1/\delta_\mu$, and the fact that the vibrational eigenvectors $b_{i\mu}$ form an orthonormal set, i.e.~$\sum_\mu b_{i\mu}b_{j\mu}=\delta_{ij}$.
When, for instance, the COM mode dominates the dynamics, all $\mu\neq \mathrm{COM}$ summands in $\varphi_{ij}$ are suppressed by $1/\delta_\mu$ and only the $\mu=\mathrm{COM}$ survives. Thus $\varphi_{ij}\equiv\varphi$ becomes independent of $i$ and $j$ and the difference between the TWA and exact solutions is simply one power of $\cos(\varphi)$. A similar reasoning applies when the detuning is tuned close to a different mode.
However, consider the limit of a large enough detuning $\delta_\mu$ such that it is approximately $\mu$-independent.\footnote{Strictly speaking one needs to go beyond the RWA in the limit $\delta\rightarrow\infty$. However, we emphasize that the arguments that follow are valid as well without the RWA.}
In this limit, $\varphi_{ij} \sim \sum_\mu b_{i\mu}b_{j\mu}/\omega_\mu$. If the frequencies $\omega_\mu$ are all of the same order of magnitude we have that $\varphi_{ii}$ is a sum of positive terms, whereas $\varphi_{i\neq j}$ is a sum of positive and negative terms that tend to cancel each other due to the orthonormality of $b_{i\mu}$.
Therefore, for large detuning the TWA solution, $\langle S^x_i \rangle_\cl = \cos(\varphi_{ii}) \langle \hat{\sigma}^x_i \rangle$, acquires an extra cosine factor with an argument $\varphi_{ii} \gg \varphi_{i\neq j}$ oscillating much faster than typical time scales. This makes the TWA prediction deviate significantly from the quantum solution already at early times. To cure the self-interaction and, more generically, to add corrections to the spin-boson TWA we follow the method proposed in \cite{PhysRevB.93.174302}, which is based on using an extended set of equations similar to a BBGKY hierarchy.


\section{BBGKY extension\label{sec:bbgky}}

In this section, we show how to add corrections to the TWA by following the method proposed in \cite{PhysRevB.93.174302}, which consists of increasing the number of classical variables and evolving them with BBGKY-type equations. For this we go back to the Wigner-Weyl representation introduced in Section~\ref{sec:twa} and work in the Heisenberg picture, where the time dependence is in the Weyl symbols of operators instead of the Wigner function. The evolution equations for the Weyl symbols can be obtained by first deriving the Heisenberg equations of motion for the operators and then applying the Weyl transformation to both sides. In this way, one obtains an infinite hierarchy of coupled equations for the Weyl symbols, analogous to the BBGKY hierarchy of equations for correlation functions. For instance, for the system considered here the evolution of $(\hat{\sigma}_i^x)_W$ will depend on $(\hat{a}_\mu \hat{\sigma}_i^y)_W$, which in turn depends on $(\hat{a}_\mu \hat{a}_\nu \hat{\sigma}_i^x)_W$ and so on. In order to truncate this infinite hierarchy of equations one may define connected Weyl symbols analogous to connected correlation functions, e.g.~$(\hat{a}_\mu \hat{\sigma}_i^y)_W \equiv (\hat{a}_\mu \hat{\sigma}_i^y)_{W,c} + (\hat{a}_\mu)_W (\hat{\sigma}_i^y)_W$. One can show that by setting all connected parts to zero, i.e.~by splitting all Weyl symbols of products of operators into products of Weyl symbols, one recovers the TWA. A natural extension of TWA is hence to avoid this splitting and take the evolution of the connected parts into account. In the following we elaborate in more depth this idea.

To simplify the notation we define Weyl symbols of products of operators as
\begin{align}
\begin{aligned}[c]
	S_{i}^{\alpha} \equiv&\, ( \hat{\sigma}_i^\alpha )_W \, , \\
	S_{ij}^{\alpha\beta} \equiv&\, ( \hat{\sigma}_i^\alpha \hat{\sigma}_j^\beta )_W \, , \\
	S_{ijk}^{\alpha\beta\gamma} \equiv&\, ( \hat{\sigma}_i^\alpha \hat{\sigma}_j^\beta \hat{\sigma}_k^\gamma )_W \, , 
\end{aligned} \qquad
\begin{aligned}[c]
	A_{\mu} \equiv&\, (\hat{a}_\mu)_W \, , \\
	A_{\mu\nu}^{00} \equiv&\, (\hat{a}_\mu \hat{a}_\nu)_W \, , \\
	A_{\mu\nu}^{10} \equiv&\, \frac{1}{2}(\{ \hat{a}_\mu^\dagger, \hat{a}_\nu \})_W \, , 
\label{eq:defs_weylsSA}
\end{aligned}
\end{align}
\begin{align}
\begin{aligned}[c]
	M_{i\mu}^\alpha \equiv&\, (\hat{\sigma}_i^\alpha \hat{a}_\mu)_W \, ,& \\
	M_{ij\mu}^{\alpha\beta} \equiv&\, (\hat{\sigma}_i^\alpha \hat{\sigma}_j^\beta \hat{a}_\mu)_W \, , \\
	M_{i\mu\nu}^{\alpha10} \equiv&\, \frac{1}{2}(\hat{\sigma}_i^\alpha \{ \hat{a}_\mu^\dagger , \hat{a}_\nu \})_W \, ,
\label{eq:defs_weylsM}
\end{aligned}
\end{align}
where $i\neq j \neq k \neq i$, all products of operators are symmetrized, and in this notation `$0(1)$' stands for $\hat{a}^{(\dagger)}$. Note that products of operators are automatically symmetrized when the operators act on different sites or modes. We define connected Weyl symbols (calligraphy letters) in the same way one would do it for expectation values, e.g.
\begin{align}
\begin{aligned}
	S_{ij}^{\alpha\beta} =&\, \cS_{ij}^{\alpha\beta} + S_i^\alpha S_j^\beta \, , \\
	A_{\mu\nu}^{10} =&\, \cA_{\mu\nu}^{10} + A_\mu^* A_\nu \, , \\
	M_{i\mu}^\alpha =&\, \cM_{i\mu}^\alpha + S_i^\alpha A_\mu \, ,
\label{eq:defs_connweyl}
\end{aligned}
\end{align}
and analogously for higher order products.
The equations of motion for the connected Weyl symbols $\cS$, $\cA$ and $\cM$ can be obtained by combining the equations for the full Weyl symbols $S$, $A$ and $M$. As explained above, the latter follow directly from the Heisenberg equations of motion for the corresponding operators. To truncate the hierarchy one needs to discard connected Weyl symbols of high orders. In this work, we take all two-point functions into account and neglect all third and higher order functions with one exception: the spin-spin-boson symbol $\cM_{ij\mu}^{\alpha\beta}$. The reason for this is that these terms turn out to be relevant in the evolution of spin-spin correlators. The resulting equations are provided in Appendix~\ref{app:bbgky_eq}.

\begin{figure*}[t!]
	\centering
	\includegraphics[width=\textwidth]{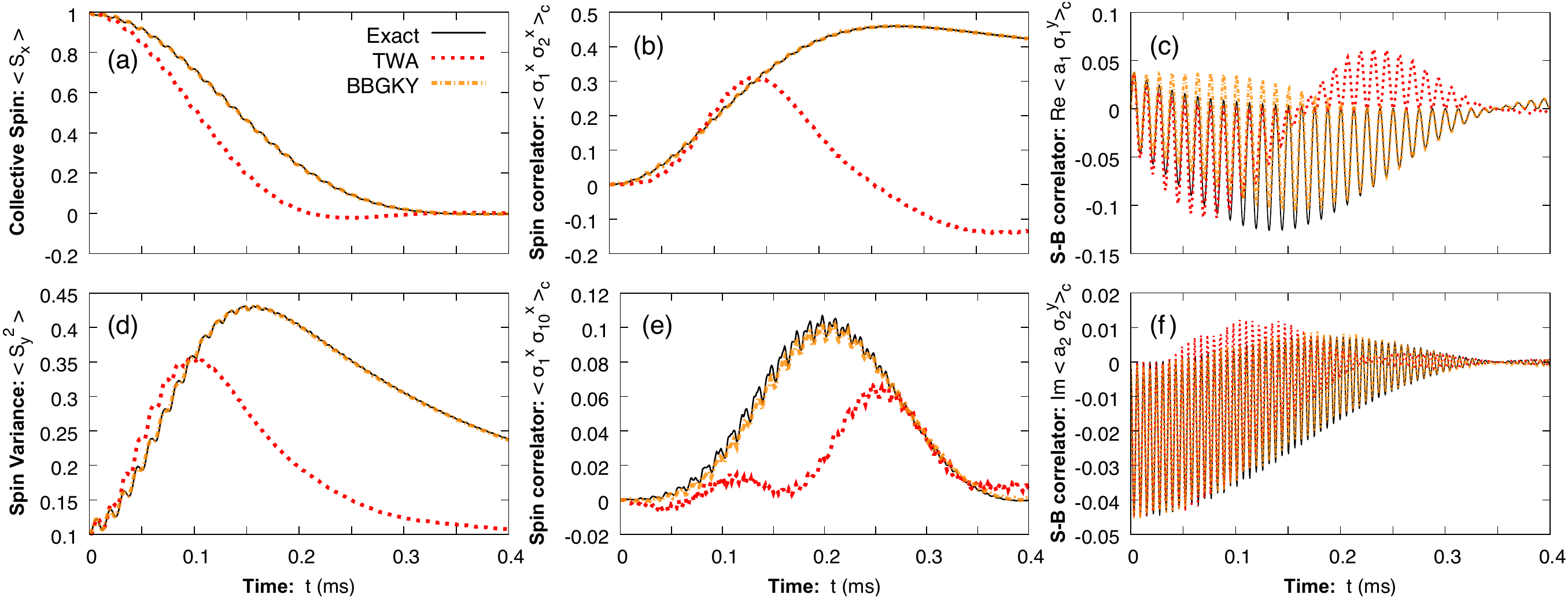}
	\caption{Comparison of collective, single spin, and mixed spin-boson observables using TWA, BBGKY and the exact solution for the many-mode case with $N_s=N_b=10$, $\delta=80\,\text{kHz}$ and $\Omega=19.42 \,\text{kHz}$. In the above panels we demonstrate how using the BBGKY extension allows one to greatly improve on the TWA in capturing the dynamics of collective spin length (a) and variance (d), spin-spin correlators (b) and (e), as well as spin-phonon correlators (c) and (f).}
\label{fig:monroe_manymode}
\end{figure*}

The success of the BBGKY hierarchy depends on the correct initialization of $\cS$, $\cA$ and $\cM$. To this end, consider the classical average of some product of classical variables, i.e.~of one-point Weyl symbols, at $t=0$. According to the Wigner-Weyl framework, this gives precisely the expectation value of the corresponding symmetrized product of operators, e.g.~$\langle \hat{\sigma}^x_i \hat{\sigma}^x_j \rangle = \int \mathrm{d}\mbf{S}_0 W(\mbf S_0) S^x_i S^x_j$, as long as one is using the correct Wigner function.
Therefore, all connected Weyl symbols are exactly zero at the beginning. This is in particular true regardless of whether one has correlated or uncorrelated initial conditions.

Finally, it is important to note that two-point connected Weyl symbols are not in one-to-one correspondence with connected correlators. To see this recall the relation between the Weyl symbols and the corresponding expectation values outlined in Section~\ref{sec:twa}. For instance, one would have
{\allowdisplaybreaks[0]
\begin{align}
	\langle \hat{\sigma}_i^\alpha \hat{\sigma}_j^\beta \rangle_{\text{c}} =&\, \langle \hat{\sigma}_i^\alpha \hat{\sigma}_j^\beta \rangle - \langle \hat{\sigma}_i^\alpha \rangle \langle \hat{\sigma}_j^\beta \rangle \nonumber\\
	=&\, \int \mathcal{D}\mbf{S}_{0} \left[ \cS_{ij}^{\alpha\beta} + S_i^\alpha S_j^\beta \right] - \int \mathcal{D}\mbf{S}_{0}\, S_i^\alpha \int \mathcal{D}\mbf{S}_{0}\, S_j^\beta \nonumber \\
	\neq&\, \int \mathcal{D}\mbf{S}_{0}\, \cS_{ij}^{\alpha\beta} \, ,
\label{eq:exp_conn_relation}
\end{align}
}%
where $\mathcal{D}\mbf{S}_{0} = \prod_k \mathrm{d}\mbf{S}_{k,0} W(\mbf{S}_{k,0})$. In other words, sampling $\cS_{ij}^{\alpha\beta}$ alone does not yield the connected part $\langle \hat{\sigma}_i^\alpha \hat{\sigma}_j^\beta \rangle_{\text{c}}$ and neglecting $\cS_{ij}^{\alpha\beta}$, as done in the usual TWA, does not necessarily make connected correlators vanish.


\section{Many modes\label{sec:manymode}}

In this section we apply the BBGKY method to the same system as in Sec~\ref{sec:singlemode} but operate in a regime where many phonon modes contribute to the dynamics. We choose the experimentally relevant parameters \cite{Richerme2014} $\delta=80\,\text{kHz}$ and $\Omega=19.42 \,\text{kHz}$ and perform the simulation using $N_b=10$ modes. This is the other regime of interest in current trapped-ion simulators and can be accessed if $\delta_\mu \gg \Omega_\mu$ for all modes of the crystal. In this regime the spin-spin interactions mediated by the phonons decay approximately as a power-law $1/r^\alpha$ with the distance, where $\alpha$ increases monotonically with the detuning. For the above parameters the range of the interaction is approximately given by $\alpha\approx0.58$ (see Appendix~\ref{app:modes}).

Figs.~\ref{fig:monroe_manymode}(a) to (f) show the evolution of a representative selection of observables: the collective spin $\langle \hat{S}_x \rangle$, the spin variance in the orthogonal direction $\langle \hat{S}_y^2 \rangle$, spin-spin correlators $\langle \hat{\sigma}_1^x \hat{\sigma}_j^x \rangle_c$ for $j=2,10$ and the spin-phonon correlator $\langle \hat{a}_\mu \hat{\sigma}_i^y \rangle_c$ for $\mu=1,2$ and $i=1,2$, respectively. As compared to the single mode case of Fig.~\ref{fig:monroe_single} the dynamics happen at a shorter time scale due to the larger value chosen for the coupling $\Omega$. The larger detuning leads however to the spin-phonon coupling being effectively weaker and hence the oscillations caused by the rotation of the phonons are not only faster but their amplitude is also smaller.
Because of this, the spin-spin and spin-phonon correlations built up are weaker as for the single mode case and partly decay at long times, see Figs.~\ref{fig:monroe_single}(b), (c), (e) and (f). For spin-spin correlations one finds, as expected, that the larger the distance between the spins the smaller the correlations are (c.f. Fig.~\ref{fig:monroe_single}(b) and (e)). Similarly, for phonons that are further away from resonance the spin-phonon entanglement created is smaller (c.f. Fig.~\ref{fig:monroe_single}(c) and (f)).

The results obtained with TWA (red) and with its BBGKY extension (orange) are shown in Fig.~\ref{fig:monroe_manymode} for comparison with the exact solution (black). As anticipated above, the TWA solution deviates from the exact one at relatively short times. The reason for this is that the effect of self-interactions becomes increasingly important as the detuning $\delta_\mu$ becomes larger (see end of Section~\ref{sec:singlemode}).
Remarkably, the corrections introduced by the BBGKY extension explained above lead to a large improvement, as seen by the agreement between the BBGKY and the exact dynamics. While some small deviations still persist (see e.g.~Fig.~\ref{fig:monroe_manymode}(c)), the error of the approximation can be further reduced by extending the hierarchy to include higher order terms, e.g.~$\cM_{i\mu\nu}^{\alpha}$. In fact, the inclusion of the third order terms $\cM_{ij\mu}^{\alpha\beta}$ proved to be crucial for obtaining a good agreement between the spin-spin correlators computed with BBGKY and the exact result, see Figs.~\ref{fig:monroe_manymode}(b) and (e).
We emphasize here that dealing with the quantum dynamics of a multi-mode spin-boson system is a highly complex task. The fact that the BBGKY method is allowing us to accurately reproduce the dynamics of this non-trivial case opens the exciting opportunity to use the method as a new tool for tackling problems that currently are intractable with other state-of-the-art numerical approaches.


\section{Role of sampling in BBGKY\label{sec:nosampling}}

Since the form of the BBGKY equations for the connected Weyl symbols is exactly the same as for the corresponding connected correlation functions one may wonder how relevant the sampling over initial conditions is. In the ideal case in which one would be able to solve the full hierarchy of equations without approximation, the sampling would not be needed since one could solve for the correlation functions directly.\footnote{One may check this in a system with a small number of $N$ spins where the hierarchy closes at order $N$.} However, when truncating the hierarchy of equations the sampling over initial conditions can lead to significant improvements, as we show in the following.

\begin{figure}[t!]
	\centering
	\includegraphics[width=\columnwidth]{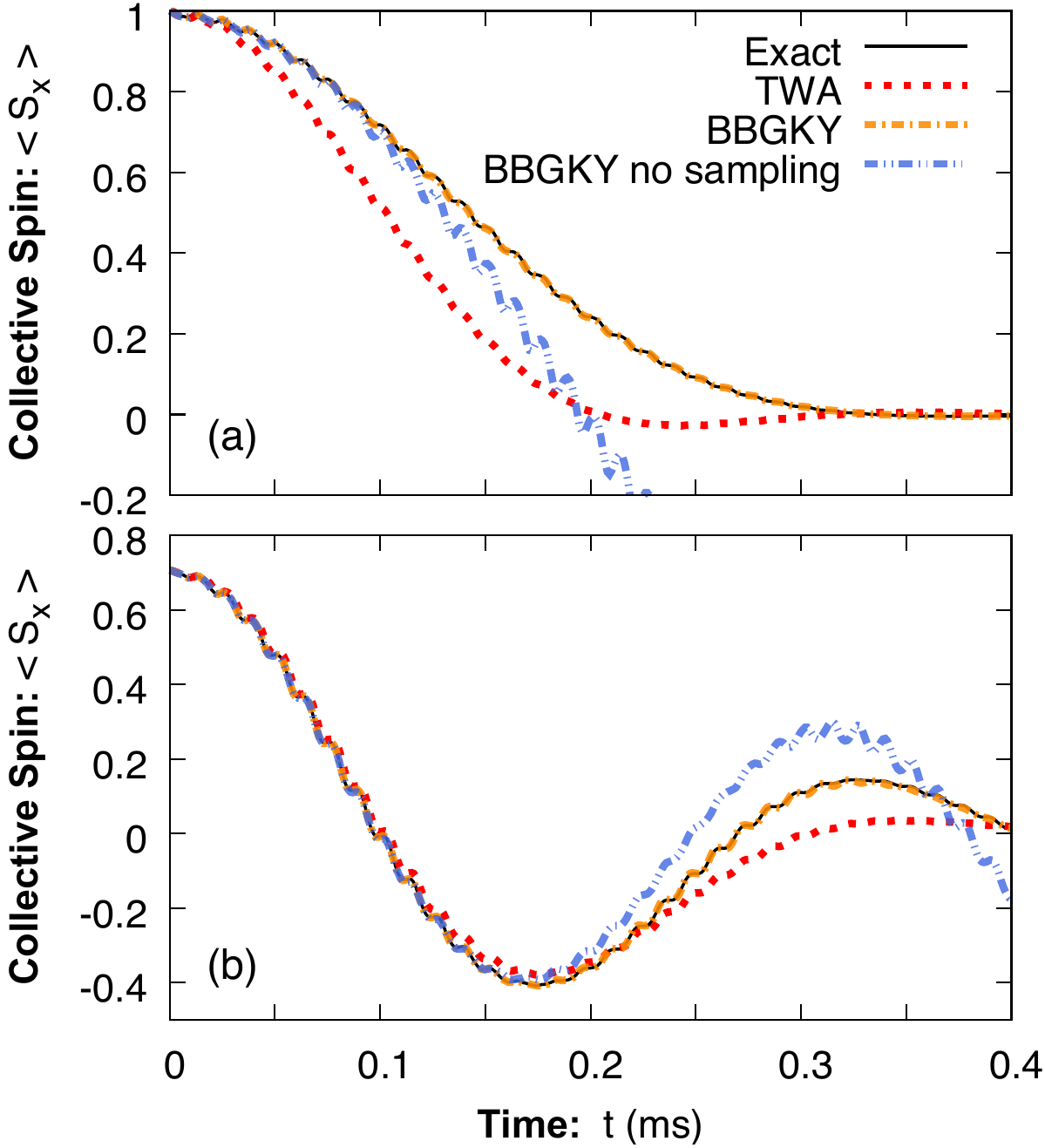}
	\caption{Evolution of the collective spin $\langle S_x \rangle$ as obtained from BBGKY with and without sampling, TWA and the exact solution for the many-mode case with $N_s=N_b=10$. Initial conditions: the bosons start in the vacuum and the spins along $(\cos(\theta/2) {|\!\! \uparrow\rangle} + \sin(\theta/2) {|\!\! \downarrow\rangle})^{\otimes N_s}$ with (a) $\theta=\pi/2$ and (b) $\theta=\pi/4$.}
\label{fig:monroe_nosampling}
\end{figure}

Solving the BBGKY equations without sampling means to initialize each correlation function to its value at initial time. For the initial condition $| \rightarrow \rangle^{\otimes N_s}\otimes | 0 \rangle^{\otimes N_b}$, the only nonzero one-point function is given by $\langle \hat{\sigma}_i^x \rangle=1$. Since the initial state is uncorrelated, all two- and three-point connected correlation functions that we take into account are initially zero except $\langle \frac{1}{2}\{ \hat{a}_\mu,\hat{a}_\mu^\dagger \} \rangle_c=1/2$. Using these initial conditions we solve the BBGKY equations for the same parameters as in Fig.~\ref{fig:monroe_manymode}. Fig.~\ref{fig:monroe_nosampling}(a) shows the evolution of the collective spin obtained from this prescription (blue line) against TWA, BBGKY with sampling and the exact solution. Although the BBGKY without sampling correctly predicts the evolution at early times, it clearly differs from the exact solution for times $t\gtrsim0.1\,\text{ms}$. 

To extend the analysis to other scenarios we consider in Fig.~\ref{fig:monroe_nosampling}(b) the state $(\cos(\pi/8) {|\!\! \uparrow\rangle} + \sin(\pi/8) {|\!\! \downarrow\rangle})^{\otimes N_s}$, for which the mean value of the spins is initially given by $\langle \hat{\boldsymbol{\sigma}}_i \rangle=(1/\sqrt{2},0,1/\sqrt{2})^T$. At early times all methods agree with the full solution. The BBGKY with sampling is however the only method that lies perfectly well on top of the exact solution over the whole time-range shown. In contrast, TWA and the BBGKY without sampling show significant deviations from the exact solution for times $t\gtrsim0.2\,\text{ms}$.
These two examples clearly show that when using BBGKY equations sampling over initial conditions can lead to a large improvement.


\section{Large system sizes\label{sec:thermal2D}}

\begin{figure*}[t!]
	\centering
	\includegraphics[width=\textwidth]{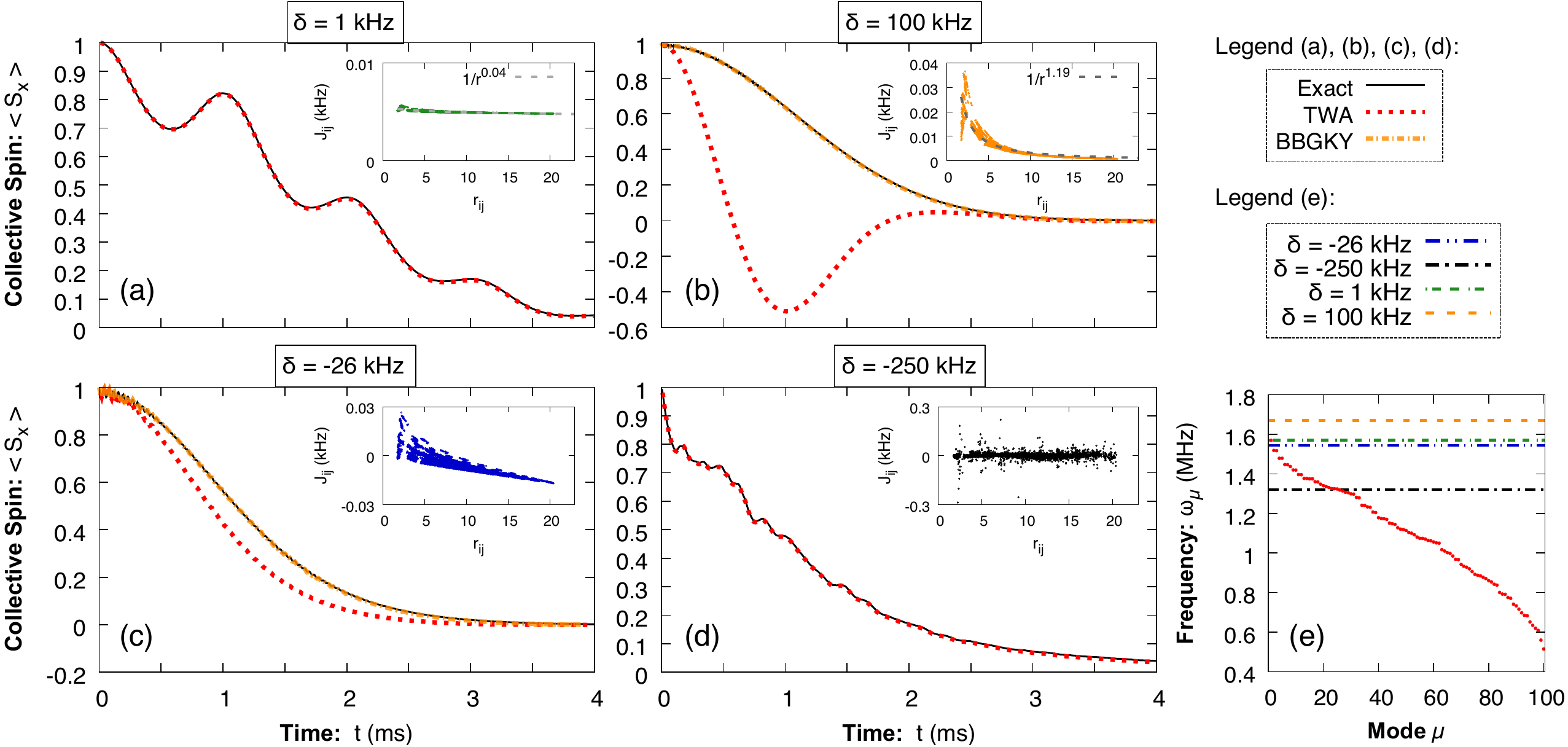}
	\caption{(a)-(d) Comparison of the collective spin $\langle S_x \rangle$ using TWA and BBGKY for thermal initial conditions in two spatial dimensions. The different plots correspond to the following values $\{\delta,\Omega\}$ of detuning and force: (a) $\{1,1\}$ kHz, (b) $\{100,12\}$ kHz, (c) $\{-26,5\}$ kHz and (d) $\{-250,2\}$ kHz. (e) Phonon frequencies for 2D. The dashed lines correspond to the position of the beatnote frequency $\omega_R$ for the different detunings used.}
\label{fig:thermal2D}
\end{figure*}

In the previous sections we have shown that the spin-boson TWA together with a systematic BBGKY expansion reproduces the dynamics of the spin-boson model (\ref{eq:HRWA}) for both the single-mode and the many-mode case. While so far we have considered small systems in one dimension, here we show that the method can easily be applied to larger systems in higher dimensions and with realistic initial conditions for the phonons.

To this end we consider a system composed of $N_s=100$ spins and $N_b=100$ phonon modes in two dimensions in a setting similar to the experiment of Ref.~\cite{Britton2012}. We initialize the spins in the fully magnetized $|\!\! \rightarrow \rangle$ state. To mimic realistic experimental conditions we let each phonon mode $\mu$ start in a thermal equilibrium state $\hat{\rho}_{th}\equiv \bigotimes_\mu e^{-\beta_\mu \hat{H}_{b,\mu}}/\Tr(e^{-\beta_\mu \hat{H}_{b,\mu}})$ with $\hat{H}_{b,\mu}=\omega_\mu \hat{n}_\mu$. We assign to each phonon initially a fixed but random mean occupancy of $\bar{n}_\mu = 5 + \eta_\mu$ where $\eta_\mu$ is a gaussian random number with $\langle \eta_\mu \rangle=0$ and $\sqrt{\langle \eta_\mu^2 \rangle} = 2 $. This sets the inverse temperature $\beta_\mu=1/T_\mu$ using (\ref{eq:nbar}).

Figs.~\ref{fig:thermal2D}(a)-(d) show the evolution of the total magnetization $\langle \hat{S}_x \rangle$ for a range of different detunings $\delta = \{ 100, 1, -26, -250 \}\,\text{kHz}$, where the force is respectively given by $\Omega = \{ 12, 1, 5, 2 \}\, \text{kHz}$. As the detuning is changed, the coupling strength of the spins to the different modes $\mu$ varies depending on how close the beatnote frequency, $\omega_R=\omega+\delta$, is to the corresponding mode frequency $\omega_\mu$. In Fig.~\ref{fig:thermal2D}(e) we show for reference the values of the different $\omega_R$ as compared to the mode frequencies. Because of this, the nature of the effective phonon-mediated Ising interaction $J_{ij}$ changes as well. In the inset of Figs.~\ref{fig:thermal2D}(a)-(d) we show $J_{ij}$ for the corresponding detuning as a function of the inter-particle distance $r_{ij}$.

For $\delta = 1\,\text{kHz}$ the spins couple predominantly to the COM mode and the Ising interaction is practically uniform ($\alpha\approx0.04$), analogous to Section~\ref{sec:singlemode}. This can be observed in the slow single-frequency oscillations on top of the decay of the magnetization in Fig.~\ref{fig:thermal2D}(a). The amplitude of these oscillations is however comparably larger due to the initial thermal occupation of the phonons. As compared to the results presented in Fig.~\ref{fig:monroe_single}, Fig.~\ref{fig:thermal2D}(a) shows that TWA (red) agrees even better with and in fact lies almost perfectly on top of the exact solution (black). The reason behind this is that, in the single mode case, the self-interaction term responsible for the deviation becomes negligible in the limit $N_s\rightarrow\infty$, as argued in Section~\ref{sec:singlemode}.

For $\delta = 100\,\text{kHz}$ we have a situation similar to Section~\ref{sec:manymode}, where many weakly coupled modes contribute to the evolution. The Ising interaction is characterized in this case by a power-law with $\alpha\approx1.19$. Due to the large detuning the coupling to the phonon modes is weaker and hence the amplitude of the oscillations is practically negligible, as shown in Fig.~\ref{fig:thermal2D}(b). Analogous to the results presented in Section~\ref{sec:manymode}, we see here that the TWA again fails to accurately describe the dynamics of the systems, whereas the BBGKY method (orange) adds the necessary corrections to make the result lie remarkably well on top of the exact solution. 

To explore other parameter regimes we consider as well negative detunings. Compared to positive detunings the landscape of Ising interactions is less trivial and can not be captured by a simple power-law \cite{Lewenstein2016} (see insets of Figs.~\ref{fig:thermal2D}(c) and (d)). For $\delta = -250\,\text{kHz}$ the beatnote frequency $\omega_R$ lies within a dense region of mode-frequencies (see Fig.~\ref{fig:thermal2D}(e)) and hence the spins couple strongly to a handful of modes. As shown in Fig.~\ref{fig:thermal2D}(d) this leads to superposed oscillations of different frequencies, whose amplitude is additionally enhanced by the initial thermal occupation of the phonons. Once again, we find that the TWA result lies almost perfectly on top of the exact solution and no BBGKY is needed. This is consistent with the argument presented in Section~\ref{sec:singlemode} that the effect of the self-interaction term in TWA is ameliorated when $\delta_\mu$ varies strongly with $\mu$.

Lastly, we consider $\delta = -26\,\text{kHz}$, which makes $\omega_R$ lie approximately equidistant from the COM and the next two phonon modes (see Fig.~\ref{fig:thermal2D}(e)). Because the detuning from the nearest modes is relatively large, the amplitude of the oscillations remains small as seen in Fig.~\ref{fig:thermal2D}(c). This choice of detuning implies as well that the coupling of the spins to the modes does not vary strongly enough with $\mu$ to make the self-interaction term negligible. Therefore, TWA shows a considerable deviation from the exact solution, which is, however, cured again using the BBGKY method.

In summary, we have shown that TWA and BBGKY can be efficiently used for large system sizes, higher dimensions and thermal initial conditions. TWA accurately describes the spin-boson dynamics when a handful of modes are dominant, whereas the BBGKY method adds the necessary corrections whenever this is not the case. Although for large system sizes the number of equations to be solved grows rapidly, we note that for the initial conditions considered in this section and for computing just the collective spin, the set of equations to be solved can be significantly reduced without further approximations, see Appendix~\ref{app:bbgky_eq}.


\section{Conclusions and Outlook\label{sec:conclusions}}

In this paper we studied the unitary dynamics of a system of spins coupled to one or many bosonic modes, where the system parameters require one to treat the spins and the bosons on equal footing. Such models are particularly relevant to experimental AMO systems, such as those realized by a 1D chain or a 2D array of trapped-ions used for quantum simulation of a variety of spin Hamiltonians. Our treatment is based on the truncated Wigner Approximation (TWA) where one samples a range of initial conditions and evolves each with classical mean-field equations. We began by adapting the TWA to spins and bosons, making use of recently developed discrete sampling methods for the spins. We further improved the TWA by introducing more classical variables and equations in a fashion similar to a BBGKY hierarchy to capture the dynamics of the higher order correlations in the system. From this point of view, TWA emerges as the lowest order approximation in the BBGKY hierarchy. We tested the convergence of the various approximations extensively by comparing to exact dynamics. We found excellent agreement for various one- and two-point functions for a large range of parameters.
Specifically, if a handful of bosonic modes dominates the dynamics, the TWA was found to give accurate results while in all remaining cases the BBGKY extension added the necessary corrections allowing the results to converge to the exact ones. We emphasize that the restriction to few modes found for TWA does not necessarily generalize to other models.
Most importantly, we demonstrated the applicability of the method to large systems ($N\sim100$) in higher dimensions (2D) with a thermal occupation of the bosonic modes, a case relevant to current experiments.

Of course better convergence always comes with a cost. The inclusion of higher orders of classical variables increases the computational time by an order $\sim N$ thus limiting the size of the systems which can be simulated efficiently. However it is worth pointing out that depending on the observable of interest one may need to go to different orders in the BBGKY approximation  which can reduce the computational complexity. The fundamental challenge, which should be addressed in the future, is the application of the BBGKY method to more generic problems that do not admit an exact solution. In this case, the truncated hierarchy of equations can become numerically unstable and a systematic evaluation of the parameter regime of validity will be necessary. Nevertheless, the numerical results presented in this work are encouraging and suggest that the BBGKY hierarchy combined with a sampling over initial conditions can become a powerful numerical method to compute any order observable at polynomial computational cost for any system size and in arbitrary dimensions. This opens a new promising path for the study of a myriad of models which cannot be simulated using other currently available methods.

\begin{acknowledgments}
	We thank Johannes Schachenmayer, Justin Bohnet, John Bollinger and J\"urgen Berges for insightful discussions and collaboration on related work. A.P.O.~thankfully acknowledges funding from the German Academic Exchange Service (DAAD) under the grant Kurzstipendien f\"ur Doktoranden (57044996). This work is supported by the DFG Collaborative Research Centre “SFB 1225 (ISOQUANT)”. We acknowledge as well support from NSF-PHY 1521080, JILA-NSF-PFC-1125844, DARPA (W911NF-16-1-0576 through ARO), MURI-AFOSR and AFOSR.
\end{acknowledgments}

\bibliography{mybibliography_ions_dTWA}


\appendix

\section*{Supplementary Material}


\section{Analytical sampling of classical equations of motion\label{app:exact_twa}}

In this section we outline how to obtain the expressions (\ref{eq:expX_twa}), (\ref{eq:expXX_twa}) and (\ref{eq:expASy_twa}) given in the main text by analytically sampling the classical equations of motion, Eq.~(\ref{eq:ceom_gen}). According to these equations, the $z$-component of the spin is a constant of motion $S_i^z(t)=S_i^z(0)\equiv z_{i,0}$. The solution for the other variables is given by
\begin{align}
\begin{aligned}
	A_\mu(t)=&\, \frac{1}{2\delta_\mu} \left( 1 - e^{-\ri\delta_\mu t} \right) \sum_i \Omega_{i\mu} z_{i,0} + A_{\mu,0} , \\
	S_i^x(t) =&\, s_\perp \sin\left( -\sgn(y_{i,0})J_i(t) + \theta_{x,i} \right) \, , \\
	S_i^y(t) =&\, s_\perp \sin\left( \sgn(x_{i,0})J_i(t) + \theta_{y,i} \right) \, ,
\label{eq:eom_sol}
\end{aligned}
\end{align}
where we defined $A_{\mu,0}\equiv A_{\mu}(0)$, $x_{i,0}\equiv S_i^x(0)$, $y_{i,0}\equiv S_i^y(0)$, $\theta_{x,i}\equiv\arcsin(x_{i,0}/s_\perp)$, $\theta_{y,i}\equiv\arcsin(y_{i,0}/s_\perp)$, $s_\perp^2=x_{i,0}^2+y_{i,0}^2$ and (c.f.~Eq.~(\ref{eq:def_varphi}))
\begin{equation}
	J_i(t) = \sum_j \varphi_{ij}(t) z_{j,0} - 2 \sum_\mu \frac{\Omega_{i\mu}}{\delta_\mu} \Im\left\{ A_{\mu,0} \left( e^{\ri\delta_\mu t} - 1 \right) \right\} \, ,
\end{equation}
Note that the signs of the angles $\theta_{x,i}$ and $\theta_{y,i}$ have to fulfil $\sgn(x_{i,0})\theta_{x,i} + \sgn(y_{i,0})\theta_{y,i}=\pi/2$.

The results (\ref{eq:expX_twa}), (\ref{eq:expXX_twa}) and (\ref{eq:expASy_twa}) are obtained by sampling the solutions (\ref{eq:eom_sol}) and products thereof over initial conditions as specified by the Wigner functions (\ref{eq:boson_samplingW}) and (\ref{eq:spin_samplingW}). We denote this by $\langle \cdot \rangle_\cl \equiv \langle \langle \cdot \rangle_{\cl,S} \rangle_{\cl,B} = \langle \langle \cdot \rangle_{\cl,B} \rangle_{\cl,S}$, where $\langle \cdot \rangle_{\cl,B(S)}$ stands for sampling over boson (spin) initial conditions. Recall that performing the spin and boson samplings independently from each other is only possible if there is no correlation between spins and bosons at $t=0$. For the specific case of $| \rightarrow \rangle^{\otimes N_s} \otimes |0\rangle^{\otimes N_b}$ as initial state the averages are given by
\begin{align}
\begin{aligned}
	\langle O \rangle_{\cl,S} =&\, \frac{1}{2^{2N_s}} \prod_i \sum_{y_{i,0}=\pm1}\, \sum_{z_{i,0}=\pm1} O \, , \\
	\langle O \rangle_{\cl,B} =&\, \prod_\mu \int d^2A_{\mu,0}\, \frac{2}{\pi}\, \exp\left\{ -2|A_{\mu,0}|^2 \right\} O \, ,
\end{aligned}
\end{align}
where $x_{i,0}=1$, $d^2A_{\mu,0}\equiv d\Re(A_{\mu,0})\,d\Im(A_{\mu,0})$ and $O$ stands for some product of classical variables.


\section{BBGKY Equations\label{app:bbgky_eq}}

In the following we list the BBGKY equations of motion used in the main text, which can be obtained as explained in Section~\ref{sec:bbgky}. The equations for the one-point functions are given by
\begin{widetext}
\begin{align}
\begin{aligned}
	\dot{A}_\mu =&\, \frac{\ri}{2} e^{-\ri\delta_\mu t} \sum_i \Omega_{i\mu} S_i^z \, ,\\
	\dot{S}_i^x =&\, 2 \sum_\mu \Omega_{i\mu}\, \Re\left[ (\cM_{i\mu}^y + S_i^y A_\mu) e^{\ri\delta_\mu t} \right] \, ,\\
	\dot{S}_i^y =&\, -2 \sum_\mu \Omega_{i\mu}\, \Re\left[ (\cM_{i\mu}^x + S_i^x A_\mu) e^{\ri\delta_\mu t} \right] \, ,
\label{eq:bbgky_one}
\end{aligned}
\end{align}
where again $\dot{S}_i^z=0$.
Note that setting $\cM\rightarrow0$ one recovers the classical equations of motion (\ref{eq:ceom_gen}). Keeping the mixed spin-boson connected Weyl symbols $\cM\neq0$ one needs to supplement these equations by
\begin{align}
\begin{aligned}
	\dot{\cM}_{i\mu}^{x} =&\, \sum_\nu \Omega_{i\nu}\, \Big\{ 2\, \cM_{i\mu}^y\, \Re\left( A_\nu e^{\ri\delta_\nu t} \right) + S_i^y \left( \cA_{\nu\mu}^{00} e^{\ri\delta_\nu t} + \cA_{\nu\mu}^{10} e^{-\ri\delta_\nu t} \right) \Big\} + \frac{\ri}{2} e^{-\ri\delta_\mu t} \Big\{ \sum_{k\neq i} \Omega_{k\mu} \cS_{ik}^{xz} - \Omega_{i\mu} S_i^z S_i^x \Big\} \, , \\
	\dot{\cM}_{i\mu}^{y} =&\, - \sum_\nu \Omega_{i\nu} \, \Big\{ 2\, \cM_{i\mu}^x\, \Re \left( A_\nu e^{\ri\delta_\nu t} \right) + S_i^x \left( \cA_{\nu\mu}^{00} e^{\ri\delta_\nu t} + \cA_{\nu\mu}^{10} e^{-\ri\delta_\nu t} \right) \Big\} + \frac{\ri}{2} e^{-\ri\delta_\mu t} \Big\{ \sum_{k\neq i} \Omega_{k\mu} \cS_{ik}^{yz} - \Omega_{i\mu} S_i^z S_i^y \Big\} \, , \\
	\dot{\cM}_{i\mu}^{z} =&\, \frac{\ri}{2} e^{-\ri\delta_\mu t} \Omega_{i\mu} (1 - S_i^z S_i^z ) \, ,
\label{eq:bbgky_mixed2}
\end{aligned}
\end{align}
where we neglected three-point Weyl symbols of the form $\cM_{i\mu\nu}^{\alpha}$. These equations couple as well to the two-point Weyl symbols $\cA_{\mu\nu}$ and $\cS_{ij}^{\alpha\beta}$. Their evolution is given by
\begin{align}
\begin{aligned}
	\dot{\cA}_{\mu\nu}^{00} =&\, \frac{\ri}{2} \sum_j \bigg[ \Omega_{j\mu} \cM_{j\nu}^z e^{-\ri\delta_\mu t} + (\mu\leftrightarrow\nu) \bigg] \, , \\
	\dot{\cA}_{\mu\nu}^{10} =&\, -\frac{\ri}{2} \sum_j \bigg[ \Omega_{j\mu} \cM_{j\nu}^z e^{\ri\delta_\mu t} - (\mu\leftrightarrow\nu,\text{c.c.}) \bigg] \, , \\
\label{eq:bbgky_boson2}
\end{aligned}
\end{align}
and
\begin{align}
\begin{aligned}
	\dot{\cS}_{ij}^{xx} =&\, 2 \sum_\mu \bigg\{ \Omega_{i\mu} \Re\Big[ \left( \cM_{ij\mu}^{yx} + \cS_{ij}^{yx} A_\mu + \cM_{j\mu}^x S_i^y \right) e^{\ri\delta_\mu t} \Big] + (i\leftrightarrow j) \bigg\} \, , \\
	\dot{\cS}_{ij}^{xy} =&\, 2 \sum_\mu \bigg\{ \Omega_{i\mu} \Re\Big[ \left( \cM_{ij\mu}^{yy} + \cS_{ij}^{yy} A_\mu + \cM_{j\mu}^y S_i^y \right) e^{\ri\delta_\mu t} \Big] - (i\leftrightarrow j, x\leftrightarrow y) \bigg\} \, , \\
	\dot{\cS}_{ij}^{xz} =&\, 2 \sum_\mu \Omega_{i\mu} \Re\Big[ \left( \cM_{ij\mu}^{yz} + \cS_{ij}^{yz} A_\mu + \cM_{j\mu}^z S_i^y \right) e^{\ri\delta_\mu t} \Big] \, , \\
	\dot{\cS}_{ij}^{yy} =&\, -2 \sum_\mu \bigg\{ \Omega_{i\mu} \Re\Big[ \left( \cM_{ij\mu}^{xy} + \cS_{ij}^{xy} A_\mu + \cM_{j\mu}^y S_i^x \right) e^{\ri\delta_\mu t} \Big] + (i\leftrightarrow j) \bigg\} \, , \\	
	\dot{\cS}_{ij}^{yz} =&\, -2 \sum_\mu \Omega_{i\mu} \Re\Big[ \left( \cM_{ij\mu}^{xz} + \cS_{ij}^{xz} A_\mu + \cM_{j\mu}^z S_i^x \right) e^{\ri\delta_\mu t} \Big] \, . \\
\label{eq:bbgky_spin2}
\end{aligned}
\end{align}
where $\dot{\cS}_{ij}^{zz}=0$, ``c.c.'' stands for complex conjugate and expressions like $(i\leftrightarrow j)$ are shorthand notation for the whole expression appearing on its left, within the same level of parenthesis, after applying the indicated substitution.
The latter equations couple to the spin-spin-boson Weyl symbols $\cM_{ij\mu}^{\alpha\beta}$, which one may approximately neglect to close the hierarchy. However, for the particular problem considered in this work we found these variables to be relevant in the dynamics of correlators such as $\langle \hat{\sigma}_i^\alpha \hat{\sigma}_j^\beta \rangle$. Therefore, we take into account the evolution equations of these spin-spin-boson variables, which are given by
\begin{align}
\begin{aligned}
	\dot{\cM}_{ij\mu}^{xx} =&\, 2 \sum_\nu \left\{ \Omega_{i\nu} \left[ \cM_{ij\mu}^{yx}\, \Re\left( A_\nu e^{\ri\delta_\nu t} \right) + \cM_{i\mu}^y\, \Re \left( \cM_{j\nu}^x e^{\ri\delta_\nu t} \right) \right] + (i\leftrightarrow j) \right\} \\
	&\, + \sum_\nu \Big\{ \Omega_{i\nu} \cS_{ij}^{yx} \left( \cA_{\mu\nu}^{00} e^{\ri\delta_\nu t} + \cA_{\mu\nu}^{01} e^{-\ri\delta_\nu t} \right) + (i\leftrightarrow j) \Big\} \\
	&\, - \frac{\ri}{2} e^{-\ri\delta_\mu t} \Big\{ \cS_{ij}^{xx} \left( \Omega_{i\mu} S_i^z + \Omega_{j\mu} S_j^z \right) + \left( \Omega_{j\mu} \cS_{ij}^{xz} S_j^x + (i\leftrightarrow j) \right) \Big\} \, , \\
	\dot{\cM}_{ij\mu}^{xy} =&\, 2 \sum_\nu \left\{ \Omega_{i\nu} \left[ \cM_{ij\mu}^{yy}\, \Re\left( A_\nu e^{\ri\delta_\nu t} \right) + \cM_{i\mu}^y\, \Re \left( \cM_{j\nu}^y e^{\ri\delta_\nu t} \right) \right] - (i\leftrightarrow j, x \leftrightarrow y) \right\} \\
	&\, + \sum_\nu \Big\{ \Omega_{i\nu} \cS_{ij}^{yy} \left( \cA_{\mu\nu}^{00} e^{\ri\delta_\nu t} + \cA_{\mu\nu}^{01} e^{-\ri\delta_\nu t} \right) - (i\leftrightarrow j,x\leftrightarrow y) \Big\} \\
	&\, - \frac{\ri}{2} e^{-\ri\delta_\mu t} \Big\{ \cS_{ij}^{xy} \left( \Omega_{i\mu} S_i^z + \Omega_{j\mu} S_j^z \right) + \left( \Omega_{j\mu} \cS_{ij}^{xz} S_j^y + (i\leftrightarrow j,x\leftrightarrow y) \right) \Big\} \, , \\
	\dot{\cM}_{ij\mu}^{xz} =&\, 2 \sum_\nu \Omega_{i\nu} \left[ \cM_{ij\mu}^{yz}\, \Re\left( A_\nu e^{\ri\delta_\nu t} \right) + \cM_{i\mu}^y\, \Re \left( \cM_{j\nu}^z e^{\ri\delta_\nu t} \right) \right]  \\
	&\, + \sum_\nu \Omega_{i\nu} \cS_{ij}^{yz} \left( \cA_{\mu\nu}^{00} e^{\ri\delta_\nu t} + \cA_{\mu\nu}^{01} e^{-\ri\delta_\nu t} \right) \\
	&\, - \frac{\ri}{2} e^{-\ri\delta_\mu t} \Big\{ \cS_{ij}^{xz} \left( \Omega_{i\mu} S_i^z + \Omega_{j\mu} S_j^z \right) + \Omega_{j\mu} \cS_{ij}^{xz} S_j^z \Big\} \, , \\
	\dot{\cM}_{ij\mu}^{yy} =&\, -2 \sum_\nu \left\{ \Omega_{i\nu} \left[ \cM_{ij\mu}^{xy}\, \Re\left( A_\nu e^{\ri\delta_\nu t} \right) + \cM_{i\mu}^x\, \Re \left( \cM_{j\nu}^y e^{\ri\delta_\nu t} \right) \right] + (i\leftrightarrow j) \right\} \\
	&\, - \sum_\nu \Big\{ \Omega_{i\nu} \cS_{ij}^{xy} \left( \cA_{\mu\nu}^{00} e^{\ri\delta_\nu t} + \cA_{\mu\nu}^{01} e^{-\ri\delta_\nu t} \right) + (i\leftrightarrow j) \Big\} \\
	&\, - \frac{\ri}{2} e^{-\ri\delta_\mu t} \Big\{ \cS_{ij}^{yy} \left( \Omega_{i\mu} S_i^z + \Omega_{j\mu} S_j^z \right) + \left( \Omega_{j\mu} \cS_{ij}^{yz} S_j^y + (i\leftrightarrow j) \right) \Big\} \, , \\
\dot{\cM}_{ij\mu}^{yz} =&\, -2 \sum_\nu \Omega_{i\nu} \left[ \cM_{ij\mu}^{xz}\, \Re\left( A_\nu e^{\ri\delta_\nu t} \right) + \cM_{i\mu}^x\, \Re \left( \cM_{j\nu}^z e^{\ri\delta_\nu t} \right) \right]  \\
	&\, - \sum_\nu \Omega_{i\nu} \cS_{ij}^{xz} \left( \cA_{\mu\nu}^{00} e^{\ri\delta_\nu t} + \cA_{\mu\nu}^{01} e^{-\ri\delta_\nu t} \right) \\
	&\, - \frac{\ri}{2} e^{-\ri\delta_\mu t} \Big\{ \cS_{ij}^{yz} \left( \Omega_{i\mu} S_i^z + \Omega_{j\mu} S_j^z \right) + \Omega_{j\mu} \cS_{ij}^{yz} S_j^z \Big\} \, . \\
\label{eq:bbgky_mixed3}
\end{aligned}
\end{align}
Here, $\dot{\cM}_{ij\mu}^{zz} = 0$ and we neglected higher order terms in order to close the hierarchy.
\end{widetext}

This set of equations becomes less involved when the spins start in the state $| \rightarrow \rangle^{\otimes N}$. In this case, $S_i^z(t)=\pm1$ and it follows from Eq.~(\ref{eq:bbgky_mixed2}) that $\cM_{i\mu}^z\equiv0$. This in turn implies together with Eq.~(\ref{eq:bbgky_boson2}) that $\cA_{\mu\nu}\equiv0$. Using this in Eqs.~(\ref{eq:bbgky_spin2}) and (\ref{eq:bbgky_mixed3}) we further obtain that $\cS_{ij}^{xz}\equiv0$, $\cS_{ij}^{yz}\equiv0$, $\cM_{ij\mu}^{xz} \equiv 0$ and $\cM_{ij\mu}^{yz} \equiv 0$. With these simplifications one is left with a reduced number of equations and variables which take computationally less effort to solve. We made use of this in the results presented in Sections~\ref{sec:singlemode} and \ref{sec:manymode}.

The system of equations can be further reduced if one is only interested in computing one-point functions $\langle \hat{\sigma}_i^\alpha \rangle$ or spin-boson correlators $\langle \hat{a}_\mu \hat{\sigma}_i^\alpha \rangle$. In such a case one may use the fact that Eqs.~(\ref{eq:bbgky_one}) and (\ref{eq:bbgky_mixed2}) constitute for the initial condition $| \rightarrow \rangle^{\otimes N}$ a closed set of equations and thus neglect all other equations. The results presented in Section~\ref{sec:thermal2D} were computed using this reduced set of equations.

Despite all these simplifcations, we emphasize that for general initial conditions, such as those considered in Section~\ref{sec:nosampling}, the full set of equations has to be solved, at this order of approximation.


\section{Rotated initial conditions\label{app:IC_TWA}}

In this section we fill some details about the discrete sampling scheme used for rotated spin initial conditions, i.e.~initial states that are not aligned with $x$, $y$ or $z$. To simplify the discussion we consider a system composed of just a single spin-1/2, $\hat{\boldsymbol{\sigma}}=(\hat{\sigma}^x,\hat{\sigma}^y,\hat{\sigma}^z)^T$. The following procedure is nevertheless also applicable to systems of many spins starting in a product state.

We start by recalling the sampling used for a spin initially in the state $| \uparrow \rangle$, namely:
\begin{equation}
	W_{|\uparrow\rangle}(\mbf{S}_{0})=\frac{1}{4}\,\delta(S^z_{0}-1) \sum_{x,y=\pm1} \delta(S^x_{0}-x) \delta(S^y_{0}-y) \, .
\label{eq:spin_samplingW_app}
\end{equation}
This sampling can be shown to fulfil
\begin{align}
\begin{aligned}
	\langle (\hat{\sigma}^\alpha)^n \rangle =&\, \int \mathcal{D}\mbf{S}_0\, (S^\alpha)^n \, , \\
	\langle \{(\hat{\sigma}^\alpha)^m,(\hat{\sigma}^\beta)^n \}_S \rangle =&\, \int \mathcal{D}\mbf{S}_0\, (S^\alpha)^m (S^\beta)^n \, , \\
	\langle \{(\hat{\sigma}^x)^m,(\hat{\sigma}^y)^n,(\hat{\sigma}^z)^l \}_S \rangle =&\, \int \mathcal{D}\mbf{S}_0\,  (S^x)^m (S^y)^n (S^z)^l \, ,
\label{eq:condition_sampling}
\end{aligned}
\end{align}
where we defined $\int \mathcal{D} \mbf{S}_0 \equiv \int d\mbf{S}_0\, W(\mbf{S}_{0})$, and the symmetric product of two and three operators as 
{\allowdisplaybreaks[0]
\begin{align}
         \{A, B\}_S\equiv &\frac 12 \left(AB+BA\right),\\
	\{ A, B, C\}_S \equiv&\, \frac{1}{6} \left( ABC+ACB+BAC \right. \nonumber \\
	&\, \left. +BCA+CAB+CBA \right) \, .
\end{align}
}
Using this we will show in the following that rotated initial states of the form
 \begin{align}
	| \uparrow_R \rangle \equiv&\, e^{-\ri\phi/2}\cos(\theta/2) | \uparrow \rangle + e^{\ri\phi/2}\sin(\theta/2) | \downarrow \rangle \, 
\label{eq:rotated_IC}
\end{align}
can be sampled using the Wigner function
\begin{equation}
	W_{|\uparrow_R\rangle}(\mbf{S}_{0})=\frac{1}{4}\,\delta(S^z_{R,0}-1) \sum_{x,y=\pm1} \delta(S^x_{R,0}-x) \delta(S^y_{R,0}-y) \, .
\label{eq:spin_samplingW_rot}
\end{equation}
Here, the subscript `$R$' denotes the rotated spin variables $\mbf S = R^T \mbf S_R$ and $R$ is a rotation matrix given by
\begin{equation}
	R = \begin{pmatrix} \cos(\phi)\cos(\theta) & \sin(\phi)\cos(\theta) & -\sin(\theta) \\ -\sin(\phi) & \cos(\phi) & 0 \\ \cos(\phi)\sin(\theta) & \sin(\phi)\sin(\theta) & \cos(\theta)  \end{pmatrix} \, 
\label{eq:rotation_matrix_pauli}
\end{equation}
with $R^T R = \mathbb{1}$.

To show the accuracy of the sampling (\ref{eq:spin_samplingW_rot}) the strategy consists in rotating the whole system to make the initial state lie along the $z$-direction, sample the initial conditions in the rotated basis and then rotate back.
We write the initial state as
\begin{align}
	| \uparrow_R \rangle \equiv&\, \hat{U}_R(\phi,\theta) | \uparrow \rangle \, ,
\label{eq:rotated_back}
\end{align}
where we defined the rotation operator $\hat{U}_R(\phi,\theta) = e^{-\ri(\phi/2)\hat{\sigma}_z} e^{-\ri(\theta/2)\hat{\sigma}_y}$. Using this we define the rotated spin operators $\hat{\sigma}^\mu_R \equiv \hat{U}_R\, \hat{\sigma}^\mu\, \hat{U}_R^\dagger$, which can also be written as $\hat{\boldsymbol{\sigma}}_R = R\, \hat{\boldsymbol{\sigma}}$.

We define rotated classical spin variables $\mbf S_R = R\, \mbf S = (x_R,y_R,z_R)^T$, which are the Weyl symbols of the rotated spin matrices, $(\hat{\sigma}^\alpha_R)_W = S^\alpha_R$. In the rotated basis, the spin points in the (rotated) $z$-direction. Therefore, if we sample the rotated spins according to (\ref{eq:spin_samplingW_rot}), i.e.~$x_{R,0}=\pm1$, $y_{R,0}=\pm1$, $z_{R,0}=1$, then (\ref{eq:condition_sampling}) will be fulfilled for $\hat{\sigma}\rightarrow\hat{\sigma}_R$ and $S\rightarrow S_R$. This justifies the sampling given in (\ref{eq:spin_samplingW_rot}).

The equations of motion for the rotated spin $\mbf S_R$ can be obtained from the Heisenberg equations for $\hat{\boldsymbol\sigma}_R$ after all products of operators have been symmetrized and simplified as explained in the main text. For the spin-$1/2$ case considered in this work, they can be obtained as well by rotating the equations for $\mbf S$ as
\begin{equation}
	\dot{\mbf S}_R = R\, \dot{\mbf S}(R^T \mbf S_R) \, .
\label{eq:rotated_EOM}
\end{equation}
Again, it is essential that in deriving the equations for $\mbf S$ all products of spin matrices have been symmetrized and simplified as, e.g., $(\hat{\sigma}^x)^3 = \hat{\sigma}^x$.

\begin{figure}[t]
	\centering
	\includegraphics[width=\columnwidth]{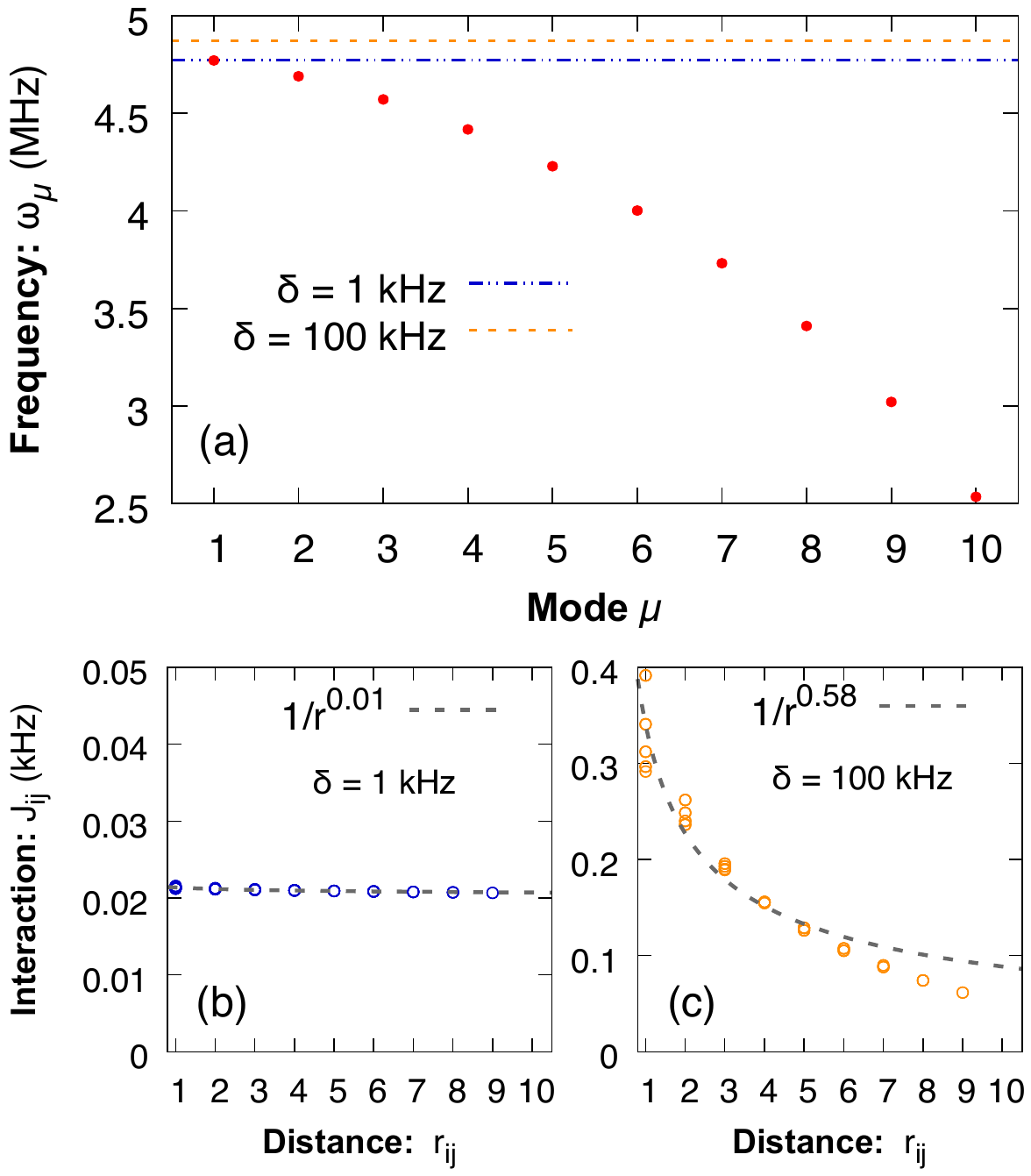}
	\caption{(a) Phonon frequencies for 1D. The dashed lines correspond to the position of the beatnote frequency $\omega_R$ for the different detunings $\delta$ used in the main text. (b), (c) Effective spin-spin interaction $J_{ij}$ as a function of interparticle distance $r_{ij}$ for the two different detunings. The grey dashed lines show the result of a fit to $1/r_{ij}^\alpha$.}
\label{fig:modesJij1D}
\end{figure}

Given (\ref{eq:spin_samplingW_rot}) and (\ref{eq:rotated_EOM}) one could in principle work in the rotated basis. However, it is usually convenient to reexpress everything in terms of the original basis. To this end, one can use Eq.~(\ref{eq:spin_samplingW_rot}) to initialize the rotated spins and then rotate back to the original basis. Rotating back (\ref{eq:rotated_EOM}) one can then evolve the spins using the original equations for $\mbf{S}$. 
In order to compute expectation values of the original spin matrices, one needs to first express the observable in terms of the rotated spin operators. Then all products have to be symmetrized and simplified. After that one substitutes $\hat{\sigma}_R^\alpha \rightarrow S_R^\alpha=\sum_\lambda R_{\alpha\lambda} S^\lambda$ and the resulting function of classical spin variables is the one to be averaged. The expectation value of, for instance, $\hat{\boldsymbol{\sigma}}$, would thus be given by
\begin{align}
	\langle \hat{\boldsymbol{\sigma}} \rangle =&\, R^T \langle \hat{\boldsymbol{\sigma}}_R \rangle \approx R^T \langle \mbf{S}_R \rangle_\cl = \langle \mbf S \rangle_\cl \, .
\label{eq:rotated_readout}
\end{align}
Here $\langle \cdot \rangle_\cl$ has to be understood as an average by sampling the rotated spins as mentioned above.

Based on the previous equality one could in principle skip the back and forth rotation and directly associate $\sigma^\alpha \leftrightarrow S^\alpha$. However, this procedure works generally only when computing observables that have been symmetrized and reduced to their simplest form. For example, if one chooses to compute $\langle(\hat{\sigma}^x)^3\rangle$ as $\langle (S^x)^3 \rangle_\cl$ instead of using $(\hat{\sigma}^x)^3 = \hat{\sigma}^x$ to compute it as $\langle S^x \rangle_\cl$ one may not obtain the correct result. The reason for this is that in general
\begin{align}
	\langle (\hat{\sigma}^\lambda)^3 \rangle =&\, \sum_{\alpha,\beta,\gamma} R^T_{\lambda\alpha} R^T_{\lambda\beta} R^T_{\lambda\gamma} \langle \hat{\sigma}^\alpha_R\, \hat{\sigma}^\beta_R\, \hat{\sigma}^\gamma_R \rangle \nonumber\\
	\stackrel{i.g.}{\neq}&\, \sum_{\alpha,\beta,\gamma} R^T_{\lambda\alpha} R^T_{\lambda\beta} R^T_{\lambda\gamma} \langle S^\alpha_R\, S^\beta_R\, S^\gamma_R \rangle_\cl = \langle (S^\lambda)^3 \rangle_\cl \, ,
\end{align}
unless the product $\hat{\sigma}^\alpha_R\, \hat{\sigma}^\beta_R\, \hat{\sigma}^\gamma_R$ happens to be automatically symmetrized. In other words, given a symmetrically ordered operator $f(\hat{\boldsymbol{\sigma}})$, we have that $(f(\hat{\boldsymbol{\sigma}}_R))_W=f(\mbf S_R)$, but in general $(f(\hat{\boldsymbol{\sigma}}))_W = (f(R^T\hat{\boldsymbol{\sigma}}_R))_W \neq f(R^T \mbf S_R) = f(\mbf S)$.


\section{Phonon modes and spin-spin interactions\label{app:modes}}

In this section we fill in some details about the phonon frequencies used for the 1D simulations of Sections~\ref{sec:singlemode}, \ref{sec:manymode} and \ref{sec:nosampling}, and compute the resulting spin-spin interactions. Fig.~\ref{fig:modesJij1D}(a) shows the frequency $\omega_\mu$ of each of the 10 phonon modes. In the same plot we give as well the value of the beatnote frequency $\omega_R = \omega + \delta$ for the different detunings $\delta$ used. For each detuning one obtains a different spin-spin interaction $J_{ij}$ mediated by the phonons, which is given by \cite{Britton2012}
\begin{equation}
	J_{ij} = \frac{1}{2} \sum_\mu \frac{\Omega_{i\mu} \Omega_{j\mu}}{\delta_\mu} \, .
\label{eq:Jij}
\end{equation}
Figs.~\ref{fig:modesJij1D}(b) and (c) show $J_{ij}$ as a function of the distance for the two different detunings $\delta=1\,\text{kHz}$ and $\delta=100\,\text{kHz}$ used. Each plot shows as well the result of a fit to a power-law decaying function $1/r_{ij}^\alpha$. We obtain $\alpha\approx0.01$ and $\alpha\approx0.58$, respectively.

\end{document}